\documentclass[aps,prl,twocolumn,superscriptaddress,longbibliography,nofootinbib,10pt]{revtex4-2}

\usepackage{amsmath,amssymb,amsfonts,amsthm}
\usepackage{braket}
\usepackage{enumitem}
\usepackage[dvipsnames]{xcolor}
\usepackage{mathtools}
\usepackage{comment}
\usepackage{algpseudocode}
\makeatletter
\newcounter{algorithm}
\renewcommand{\thealgorithm}{\arabic{algorithm}}
\def\fps@algorithm{tbp}\def\ftype@algorithm{4}\def\ext@algorithm{loa}
\def\fnum@algorithm{Algorithm~\thealgorithm}

\newcounter{suppsec}
\newcounter{suppsub}[suppsec]

\makeatother
\usepackage{hyperref}
\usepackage{booktabs}
\usepackage[section]{placeins}
\usepackage{multirow}
\usepackage{listings}
\usepackage{framed}
\usepackage{graphicx}
\usepackage{tikz}
\usetikzlibrary{positioning,arrows.meta,decorations.pathreplacing,calc}
\lstdefinestyle{spec}{
  basicstyle=\footnotesize\ttfamily,
  breaklines=true,
  breakatwhitespace=false,
  columns=fullflexible,
  keepspaces=true,
  showstringspaces=false,
  frame=single,
  framesep=4pt,
  xleftmargin=2pt,
}
\newcommand{\Ftwo}{\mathbb{F}_2}
\newcommand{\code}[3]{[[#1,\,#2,\,#3]]}

\begin{document}

\title{Large-Language-Model Discovery of Quantum LDPC Codes through Structured Concept Evolution}

\author{Zidu Liu}
\email{zidu.liu@mpl.mpg.de}
\affiliation{Max Planck Institute for the Science of Light, 91058 Erlangen, Germany}

\author{Florian Marquardt}
\email{florian.marquardt@mpl.mpg.de}
\affiliation{Max Planck Institute for the Science of Light, 91058 Erlangen, Germany}
\affiliation{Department of Physics, Friedrich-Alexander University Erlangen-N\"urnberg, 91058 Erlangen, Germany}

\date{\today}

\begin{abstract}
Quantum low-density parity-check codes promise low-overhead error correction, but high-performing constructions remain hard to find. Here we introduce structured concept evolution, in which a large language model (LLM) evolves construction programs and their internal search strategies by modifying the finite group, protograph shapes or entry-assignment rule. We search for codes with fewer than $600$ data qubits and check weight at most $12$, optimizing $kd^2/n$ or the circuit-level logical error rate. Our method improves on published lifted-product benchmarks in selected parameter regimes, with exactly certified distances for the leading results. Circuit-level optimization typically yields lower logical error rates than $kd^2/n$ optimization, often despite smaller code distances. With the same evolution framework and step budget, the best LLM outperforms all random-mutation and hill-climbing runs on both objectives. Our results point to LLM-driven program search as an approach to designing general quantum LDPC codes.
\end{abstract}

\maketitle

\section{Introduction}

Quantum computers promise computational advantages for problems ranging
from quantum simulation~\cite{daley2022practical} and chemistry~\cite{mcardle2020} to optimization and machine learning~\cite{farhi2014,Huang2022Quantum}, but realizing these advantages at scale requires quantum error correction, which encodes logical
information redundantly into many physical qubits and repeatedly extracts
syndromes from local parity checks~\cite{Shor1995Scheme,
Gottesman1997Stabilizer,Terhal2015Quantum}.
The surface code has been the standard choice for this purpose owing to its
planar geometry and high
threshold~\cite{Kitaev2003Faulttolerant,Fowler2012Surface},
and recent experiments demonstrate repeated error correction and
logical-error suppression on superconducting
processors~\cite{GoogleQuantumAI2023Suppressing,Acharya2025}.
Yet it encodes a single logical qubit in $\Theta(d^2)$ physical
qubits, a vanishing rate shared by every two-dimensional geometrically
local code~\cite{bravyi2010tradeoffs}, so its overhead remains the
central obstacle to scaling. This obstacle is now practical rather than
asymptotic: neutral-atom~\cite{manetsch2025tweezer} and trapped-ion~\cite{guo2024site} platforms have reached or approached the thousand-qubit regime, so the number of logical
qubits on a device is set directly by the rate of the underlying code.
Constant-rate codes enable fault-tolerant computation with constant
overhead~\cite{gottesman2014constant}. At the error rates and code sizes relevant to near-term hardware they reduce the cost per logical qubit by roughly an order of magnitude relative to the surface
code~\cite{bravyi2024}. Logical-qubit experiments on reconfigurable atom arrays~\cite{Bluvstein2024Logical} and superconducting
hardware~\cite{wang2026demonstration}, together with full architectural proposals for neutral-atom platforms~\cite{xu2024constant,cain2026shor}, make such codes concrete
architectural targets. This motivates the search for quantum
low-density parity-check (qLDPC) codes, which combine sparse parity
checks with finite encoding rate and growing
distance~\cite{tillich2014,Panteleev2021Degenerate,Panteleev2022Asymptotically,breuckmann2021,
Leverrier2022Quantum}.

The central difficulty is that high-performing qLDPC codes are
hard to find. The known families are built from algebraic product
constructions: hypergraph product~\cite{tillich2014}, generalized-bicycle (GB)
and related quasi-cyclic codes~\cite{mackay2004sparse,kovalev2013quantum}, including the bivariate-bicycle (BB)
codes~\cite{bravyi2024} that descend directly from them, lifted product~\cite{Panteleev2021Degenerate,panteleev2022almost}, balanced
product~\cite{breuckmann2021balanced}, fibre-bundle
codes~\cite{hastings2021fiber}, quantum Tanner codes and other
asymptotically good
constructions~\cite{Leverrier2022Quantum,Panteleev2022Asymptotically,dinur2023good}, each parametrized by a small set of group-algebra
elements or protograph polynomials. While these frameworks guarantee the CSS
commutation structure~\cite{Calderbank1996Good,Steane1996Error} by construction, the resulting performance depends on the specific group-algebra elements chosen, in a way that admits no closed-form optimum. The viable region is a vanishingly small subset of
an exponentially large discrete search space, and locating it has so far
relied largely on expert intuition, exhaustive enumeration over small
instances, or random sampling.

\begin{figure*}[t]
  \centering
  \includegraphics[width=\textwidth]{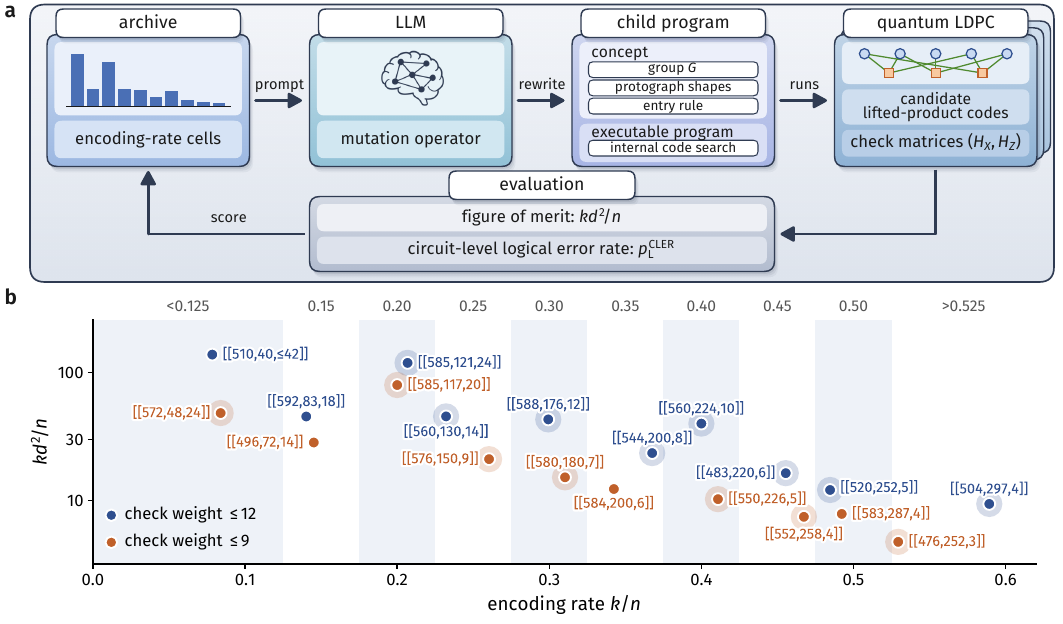}
\caption{\textbf{Structured concept evolution for quantum LDPC code discovery.} \textbf{a},~Evolutionary loop. The LLM receives a prompt based on the archive of previous solutions and rewrites the parent into a child program. The prompt contains the parent construction program, the codes it generated and their evaluation, the current search status, the program lineage and an instruction that names the part of the program to change (Methods).  Each program pairs a concept specification with an executable part written in python that may perform its own internal search to generate a batch of candidate codes. A fixed external evaluator calls the program to produce the check matrices and scores it using the figure of merit (FOM) $kd^2/n$ or the circuit-level logical error rate (CLER) $p_{\mathrm L}^{\mathrm{CLER}}$, depending on the objective we provide at the start of the entire search. The program and its score enter a quality-diversity archive that keeps the best programs of ten encoding-rate cells, together with the codes of earlier programs. \textbf{b},~Representative codes obtained through SCE for the FOM objective, one per encoding-rate cell, with check weight at most $12$ (blue) and at most $9$ (orange), shown as $kd^2/n$ against the encoding rate $k/n$. The orange code at rate $0.2$, \code{585}{117}{20}, comes from a refinement of a program from the check-weight-$9$ run (Supplemental Material~S4). Halos mark codes over non-abelian groups.}
  \label{fig:pipeline}
\end{figure*}
Machine-learning approaches have begun to address this as a combinatorial
design problem. Reinforcement learning has been used to discover
error-correction strategies, adapt codes to hardware-specific noise, and
co-optimize stabilizer codes with their encoding
circuits~\cite{fosel2018,nautrup2019,olle2024}, as well as to discover fault-tolerant state-preparation circuits under hardware constraints~\cite{zen2025quantum}, scale
automated circuit discovery~\cite{olle2025} and reduce stabilizer
measurement weight~\cite{he2025}, but these methods search over explicit, finite-size stabilizer or
circuit representations and do not yield the structured, scalable code
families required at large block length. Here we take a different route, drawing on recent advances in large language models (LLMs). In the quantum domain, LLMs have been used for scientific exploration and experiment design~\cite{Nagele2025Agentic, Arlt2026Meta}, and within multi-agent workflows for constructing nonadditive codes with prescribed transversal diagonal gates~\cite{He2025CoDesigning}. More recently, LLM-guided evolutionary search, with the LLM acting as the mutation
operator~\cite{Lehman2022Evolution}, has emerged as a powerful tool for
open-ended mathematical and algorithmic
discovery, as demonstrated in FunSearch~\cite{romeraparedes2024} and
AlphaEvolve~\cite{alphaevolve2025}.  In work contemporaneous with our initial arXiv posting, LLM-driven search was also applied to bivariate-bicycle codes~\cite{cruzbenito2026llm}. Subsequent work extended our approach to coset-orbit balanced-product codes~\cite{QianLi2026} and applied an agentic workflow with circuit-level feedback to qLDPC codes~\cite{OmniQEC2026}.

In this work, we introduce structured concept evolution (SCE), an evolutionary framework for the discovery of quantum LDPC codes. To our knowledge, SCE is the first systematic LLM-driven search over lifted-product codes, which are one of the most powerful general constructions for qLDPC codes. SCE jointly explores group-algebra entry rules, protograph shapes and general finite groups, including non-abelian ones. Each evolving individual pairs a structured concept specification with an executable program. The specification identifies a finite group, protograph shapes and an entry-assignment rule, which the program implements to generate a batch of candidate codes. The program may combine algebraic constraints, sampling and its own screening and optimization procedures. The mutation hierarchy incorporates domain knowledge by organizing program evolution around these three design choices, allowing the LLM to change both the algebraic setting and the strategy used to search for codes within it. The lifted-product construction ensures CSS commutation throughout the search. A fixed external evaluator assesses the generated codes, and a quality-diversity archive~\cite{mouret2015illuminating} retains high-scoring programs across different encoding-rate cells (Fig.~\ref{fig:pipeline}a). We run SCE separately for two objectives: the figure of merit (FOM) $kd^2/n$ and the circuit-level logical error rate in a syndrome-extraction memory experiment. In the $k/n\approx0.2$ cell, the FOM search finds \code{585}{121}{24} at check weight $12$, with $kd^2/n\approx119$. Further optimization of a program from the check-weight-$9$ run yields \code{585}{117}{20}, with $kd^2/n=80.0$. Both distances are certified exactly. The two objectives favour different constructions, and larger distances do not necessarily lead to lower logical error rates in the simulated memory experiment.

\section{Results}

\textbf{Lifted-product codes.}
We search over lifted-product (LP) CSS codes~\cite{Panteleev2021Degenerate,panteleev2022almost}, an algebraic generalization of hypergraph-product (HGP) codes that includes the first asymptotically good qLDPC families~\cite{Panteleev2022Asymptotically}.
The construction is specified by two group-algebra matrices,
$A\in\Ftwo[G]^{m_A\times n_A}$ and
$B\in\Ftwo[G]^{m_B\times n_B}$, where $G$ is a finite group of order $q=|G|$.
Let $\widetilde A$ and $\widetilde B$ denote the binary block matrices obtained by applying the left and right regular actions, respectively, to their entries.
Each block has size $q\times q$.
To distinguish the lifted product from an ordinary Kronecker product of the expanded binary matrices, define the product on protograph indices by
$(U\otimes_{\mathrm p}V)_{(i,k),(j,\ell)}=U_{ij}V_{k\ell}$,
where $U_{ij}$ and $V_{k\ell}$ are $q\times q$ blocks, and $I_r$ denotes an $r\times r$ block identity with diagonal blocks $I_q$.
The binary CSS check matrices are then
\begin{align}
H_X&=[\,\widetilde A\otimes_{\mathrm p}I_{n_B}
       \mid I_{m_A}\otimes_{\mathrm p}\widetilde B^{\mathsf T}\,],
       \nonumber\\
H_Z&=[\,I_{n_A}\otimes_{\mathrm p}\widetilde B
       \mid \widetilde A^{\mathsf T}\otimes_{\mathrm p}I_{m_B}\,].
\label{eq:lp}
\end{align}
The left and right regular actions commute, including after transposition of either action, so $H_XH_Z^{\mathsf T}=0$ for any finite group $G$~\cite{Panteleev2022Asymptotically}.
The number of physical qubits is
$n=fq$, with $f=n_An_B+m_Am_B$. Thus, for fixed protograph shapes, the block length is linear in $q$. An $X$-check weight is the sum of the support sizes in one row of $A$ and one column of $B$. A $Z$-check weight is defined analogously by one row of $B$ and one column of $A$.

SCE evolves programs that construct LP codes. At each step, the LLM modifies a parent program selected from the archive, and the child program generates a batch of $60$ candidate codes using its specified finite group, protograph shapes and entry-assignment rule (Fig.~\ref{fig:pipeline}a). The prompt specifies one of three mutation levels: level 1 modifies the entry-assignment rule, level 2 the protograph shapes, and level 3 the finite group (Methods). A fixed external evaluator builds the parity-check matrices $(H_X,H_Z)$ of the generated codes, ranks admissible candidates using a fast FOM estimate, and evaluates the top candidates using either the FOM or the CLER objective, defined below. The archive groups programs into ten encoding-rate cells to retain high-scoring programs across different encoding rates.  The run score is the sum of the best program scores over the ten cells. In the searches conducted here, we constrain every entry of $A$ and $B$ to be zero or a sum of at most five distinct group elements, every check weight to be at most $12$, and the block length to satisfy $100\le n<600$. Additional searches impose the check-weight cap of $9$ (Supplemental Material~S4).

\textbf{Objective.}
Code design involves several criteria including the block length $n$, the encoding rate $k/n$, the distance $d$, and the logical error rate, which play different roles in our searches. Block length and check weight are hard constraints, encoding rate is the archive coordinate. We perform separate runs that optimize alternative objectives. The figure of merit (FOM) $kd^2/n$ depends only on the code parameters and rewards codes that encode many logical qubits with a larger distance relative to their block length. The circuit-level logical error rate (CLER) instead is measured in a memory experiment that includes the syndrome-extraction circuit, circuit-level noise and decoder. It rewards a low failure rate under a prescribed  experimental setup (Methods).

Distance estimation could be a computational bottleneck for the FOM objective. An exact computation is far too slow for the loop. We therefore develop a GPU implementation of a syndrome-decoder search~\cite{webster2026} for light logical operators (Methods), which returns an upper bound on the distance within a fixed time budget.  We found that our distance estimation provides reasonably reliable estimates within a relatively short computation time (see the comparison with the exact distances in Supplemental Material~S5). For both objectives, each distinct candidate first undergoes a two-second GPU search for a light non-trivial logical operator, yielding a distance upper bound $d_f$. We select the $M=5$ candidates with the largest $kd_f^2/n$ for further evaluation. For the FOM objective, each selected candidate receives a thirty-second search, and $d_l$ is the smaller of the two estimates.
We define the program score as the average
\begin{equation}
\mathcal{S}_{\mathrm{FOM}}=\frac{1}{M}\sum_{i=1}^{M}
\frac{k_i d_{l,i}^{2}}{n_i}.
\label{eq:score}
\end{equation}
Missing terms contribute zero if fewer than $M$ distinct candidates are measured. Averaging over five outputs rewards a program that produces several strong candidates, rather than scoring it solely by its best instance. We note both $d_f$ and $d_l$ are search upper bounds used for ranking. Distances reported for the main-model codes are re-examined after evolution by longer searches and exact certification (Supplemental Material S5). For the CLER objective, the same $M$ candidates, selected by the two-second distance search, undergo a memory experiment with $R=6$ syndrome-extraction rounds at $p=0.002$. The memory experiment is run separately in the $X$ basis and in the $Z$ basis, each with up to $2\times10^{5}$ shots or $100$ failures. For each basis we take the upper end of the two-sided 95\% Clopper--Pearson interval of the block failure probability of code $i$, and $\bar P_i$ is the larger of the two upper ends. The score is defined as
\begin{equation}
\mathcal{S}_{\mathrm{CLER}}=-\frac{1}{M}\sum_{i=1}^{M}\log_{10}\!\left[\frac{1-(1-\bar P_i)^{1/R}}{k_i}\right].
\label{eq:plscore}
\end{equation}
 We use an effective per-round block failure rate normalized by the number of logical qubits as the circuit-level objective. Under an independent logical-failure model, it approximates the per-logical-qubit failure rate when the block failure probability is small. All generated codes are archived, and a candidate that matches an archived code, exactly or up to a relabelling of qubits and checks, receives a score of zero.

\textbf{Discovered codes.}
\label{sec:codes}
Fig.~\ref{fig:pipeline}b shows the FOM code per cell for check weight at most $12$ or $9$. Here we list the elite codes of both objectives with check weight at most $12$ in Table~\ref{tab:codes} for the seven encoding-rate cells below $k/n=0.425$. The protographs of the codes in Table~\ref{tab:codes} are given in Supplemental Material~S7. We run SCE with gpt-6.1-sol at high reasoning effort for $500$ steps, each of which generates $60$ candidates. The two FOM elite codes of the lowest-rate cell $k/n<0.125$ are two-block codes~\cite{lin2024quantum}, lifted products of two $1\times 1$ protographs with five polynomial terms per entry and check weight $10$. One is a bivariate-bicycle (BB) code over $C_{17}\times C_{17}$, \code{578}{32}{$\le$46}, and the other a generalized-bicycle (GB) code over $C_{255}$, \code{510}{40}{$\le$42}. For these two codes, the exact enumeration of low-weight logical operators up to the distance (Supplemental Material~S5) is out of reach, and we report the smaller of two upper bounds on the distance. One comes from our own improved efficient distance search algorithm (see Methods) with a  time budget of $7200$\,s. The other comes from the QDistRnd package~\cite{Pryadko2022}, which searches for light logical operators over $10^9$ random information sets per logical type. In every other cell, the distances are certified exactly by an irreducible-cluster enumeration~\cite{dumer2017}, following the presentation in Appendix~B of ref.~\cite{lee2026pp} (Supplemental Material~S5).

The rate-$0.2$ elite code \code{585}{121}{24} reaches $k d^2/n=119$ at check weight $12$.
It is a lifted product of two $1\times2$ protographs over a non-abelian metacyclic group. The codes of refs.~\cite{mitten2026,hong2026rate} use the same protograph shapes.
The CLER elite codes use a group different from the FOM elite codes in every encoding-rate cell, and a cyclic group in the cells at $0.20$, $0.25$, and $0.35$, where the FOM elite codes are on non-abelian groups. When compared against the FOM code within the same encoding-rate cell, the CLER elite codes underperform the FOM elite codes in $kd^2/n$ in every cell. To the best of our knowledge, no published lifted-product code below $600$ qubits with check weight at most $12$ reaches the FOM of the rate-$0.2$ elite code  \code{585}{121}{24}. A search with check weight at most $9$ in the same block-length range is presented in Supplemental Material~S4. With the same evolutionary framework and $500$ steps, the best code it finds across the ten encoding-rate cells is \code{585}{119}{18}, with $kd^2/n=65.9$. Further refinement of the program that generated this code yields \code{585}{117}{20}, with an exactly certified distance and $kd^2/n=80.0$. In this refinement, every step instructs the model to change only the rule that assigns the entries and to keep the group and the protograph shapes. The fifth child generates codes with these parameters in two of three independent runs. This result illustrates how the mutation hierarchy supports both exploration across algebraic settings and further optimization within a promising setting through refinement of the entry-assignment rule. Among the published weight-$9$ lifted-product codes at rate $0.2$~\cite{mitten2026,hong2026rate}, those with fewer than $600$ qubits have $d\le18$, and the shortest ones with a distance upper bound of $20$ are \code{625}{125}{$\le$20}~\cite{hong2026rate} and \code{630}{126}{$\le$20}~\cite{mitten2026}. Recently proposed pair-partition codes~\cite{lee2026pp} in the cells at $0.25$, $0.35$ and $0.40$ and GALA codes~\cite{yang2026gala} in the cell at $0.50$ reach a larger $kd^2/n$. We note that both constructions lie outside our search space, and future work based on our SCE approach can  be expanded to widen that space.

Different objectives lead not only to different code parameters but also to different construction strategies. For instance, in the $k/n<0.125$ rate cell, the FOM program that generates the GB code \code{510}{40}{$\leq$ 42} specifies common roots of its two entry polynomials to constrain the dimension, consistent with the dimension formula for GB codes~\cite{Panteleev2021Degenerate,lin2024quantum}. It ranks candidates using a cascade of lightweight searches for short logical operators. All candidates receive a short search, while the most promising receive longer searches that also consider $C_3$-symmetric logical operators. The CLER program of the same cell instead fills each entry of $A$ and $B$ with two elements of $\mathrm{Dic}_{11}$, which gives every check weight $10$.  It samples random fillings according to program-defined rules and screens them by dimension, classical minimum distance, and short-cycle counts. The remaining fillings are ranked by classical distance estimates and trapping-set scores~\cite{raveendran2021trapping}. The design statements for both codes appear verbatim in Supplemental Material~S6, and the full programs are available in the repository~\cite{repo}.

\begin{table}[h]
\centering
\caption{\textbf{Representative elite codes at different encoding rates.} ``FOM'' and ``CLER'' denote codes obtained with the figure-of-merit and circuit-level logical-error-rate objectives, respectively. The columns report the encoding rate, the group, the protograph shapes $A\times B$, the code parameters $[[n,k,d]]$, the check weight $w$, the number of CNOT gates per round of syndrome extraction (CNOTs), equal to the number of nonzero entries of $H_X$ and $H_Z$ together, and the number of CNOT layers per round in the CP-SAT schedule used for the circuit-level simulations (Depth). The distances are certified exactly, except for the two lowest-rate codes, for which only upper bounds are reported ($\dagger$).}
\label{tab:codes}
\setlength{\tabcolsep}{4pt}
\resizebox{\columnwidth}{!}{%
\begin{tabular}{llllccrr}
\toprule
$k/n$ & & Group & $A\times B$ & $[[n,k,d]]$ & $w$ & CNOTs & Depth \\
\midrule
$<0.125$ & FOM & $C_{255}$ & $(1,1)\times(1,1)$ & \code{510}{40}{$\le$42$^{\dagger}$} & 10 & 5100 & 15 \\
 & FOM & $C_{17}\times C_{17}$ & $(1,1)\times(1,1)$ & \code{578}{32}{$\le$46$^{\dagger}$} & 10 & 5780 & 15 \\
 & CLER & $\mathrm{Dic}_{11}$ & $(2,3)\times(2,3)$ & \code{572}{67}{22} & 10 & 5280 & 12 \\
\addlinespace[1.5pt]
0.125--0.175 & FOM & $C_{37}$ & $(2,3)\times(2,4)$ & \code{592}{83}{18} & 12 & 5624 & 14 \\
 & CLER & $\mathrm{Dic}_9$ & $(2,3)\times(2,4)$ & \code{576}{82}{15} & 11 & 5112 & 13 \\
\addlinespace[1.5pt]
0.175--0.225 & FOM & $C_{13}\rtimes C_9$ & $(1,2)\times(1,2)$ & \code{585}{121}{24} & 12 & 5616 & 16 \\
 & CLER & $C_{29}$ & $(2,4)\times(2,4)$ & \code{580}{116}{12} & 10 & 4350 & 13 \\
\addlinespace[1.5pt]
0.225--0.275 & FOM & $\mathrm{Dic}_7$ & $(2,4)\times(2,4)$ & \code{560}{130}{14} & 12 & 5376 & 16 \\
 & CLER & $C_{29}$ & $(2,4)\times(2,4)$ & \code{580}{132}{14} & 12 & 5568 & 16 \\
\addlinespace[1.5pt]
0.275--0.325 & FOM & $\mathrm{Dic}_{21}$ & $(1,2)\times(1,3)$ & \code{588}{176}{12} & 12 & 4284 & 15 \\
 & CLER & $C_2\times(C_7\rtimes C_3)$ & $(1,2)\times(2,6)$ & \code{588}{176}{10} & 12 & 4284 & 15 \\
\addlinespace[1.5pt]
0.325--0.375 & FOM & $Q_{16}$ & $(2,5)\times(2,6)$ & \code{544}{200}{8} & 12 & 3936 & 17 \\
 & CLER & $C_{13}$ & $(2,5)\times(3,8)$ & \code{598}{195}{7} & 12 & 4238 & 16 \\
\addlinespace[1.5pt]
0.375--0.425 & FOM & $\mathrm{AGL}(1,8)$ & $(1,3)\times(1,3)$ & \code{560}{224}{10} & 12 & 4032 & 18 \\
 & CLER & $C_{11}\rtimes C_5$ & $(1,3)\times(1,3)$ & \code{550}{220}{7} & 11 & 3520 & 16 \\
\bottomrule
\end{tabular}}
\end{table}

\textbf{Circuit-level performance.}
\begin{figure}[t]
  \centering
  \includegraphics[width=\columnwidth]{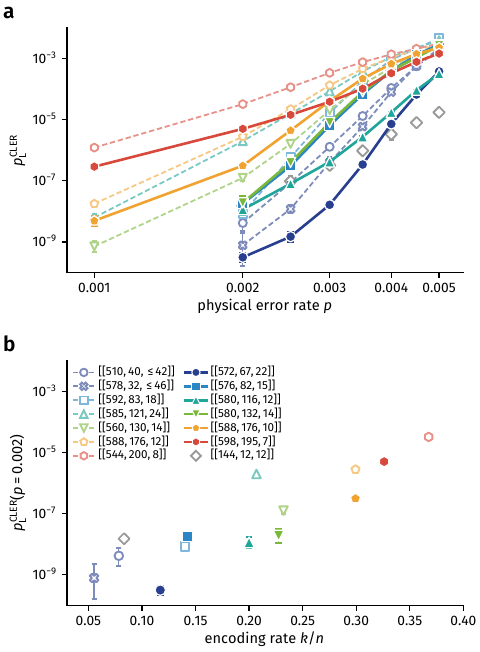}
  \caption{\textbf{Circuit-level performance.}  \textbf{a},~Effective per-round block failure rate normalized by the number of logical qubits, $p_{\mathrm L}^{\mathrm{CLER}}=[1-(1-P)^{1/R}]/k$ with $P$ the block failure probability of a basis and $R=6$, against the physical error rate $p$ for the FOM elite codes (open markers, dashed lines) and the CLER elite codes (filled markers, solid lines) of the six encoding-rate cells below $k/n=0.4$. Each point is the larger of the $X$- and $Z$-basis values, and error bars are 95\% Clopper--Pearson intervals computed from the numbers of failures and shots in that basis. Grey diamonds show the bivariate-bicycle code \code{144}{12}{12} with the published schedule of ref.~\cite{bravyi2024}, the same noise model and the same decoder.
  \textbf{b},~The same codes at $p=0.002$ against their encoding rate. 
 }
  \label{fig:sweep}
\end{figure}
To compare the two objectives at the circuit level, we measure the logical error rate of the FOM elite codes and of the CLER elite codes at different encoding rates in a memory experiment of six rounds of syndrome extraction.
Each stabilizer check is measured using a single ancilla qubit. The CNOT gates between ancilla and data qubits are scheduled in discrete time steps, with each qubit participating in at most one gate per time step, as in the edge-colouring construction of ref.~\cite{tremblay2022thin}. We use the constraint-programming solver CP-SAT to search for the schedule of smallest depth we can find, subject to the ordering condition of ref.~\cite{zhang2026asc} that ensures the circuit measures the intended stabilizers. The noise is circuit-level depolarizing noise on CNOTs, resets and measurements without idling noise, and the syndromes are decoded by relay-BP~\cite{muller2025relaybp} with post-processing~\cite{hillmann2024lsd} (see details of the circuit schedule and the decoder in Methods).

We implement the protocol with $6$ rounds and sweep the physical error rate $p$ from $0.001$ to $0.005$. We sample the noisy circuits with 
Stim~\cite{gidney2021stim}. At each $p$, we sample both bases until $1000$ failures are recorded or a shot budget is exhausted (Methods). The results are shown in Fig.~\ref{fig:sweep}. The CLER elite codes typically have lower logical error rates than their FOM counterparts at comparable encoding rates, not only at the noise strength used in the search. We remark that the advantage does not come from larger distance, since among the codes with certified distances, the CLER elites have distances no larger than those of their FOM counterparts. The rate-$0.2$ CLER elite code \code{580}{116}{12} outperforms its FOM counterpart \code{585}{121}{24} with half its distance. In the  $0.30$ cell, \code{588}{176}{12} and \code{588}{176}{10} have the same block length $n$, encoding-rate, check weight, CNOT count and circuit depth, yet the code with the smaller distance has the lower logical error rate. The reference BB code \code{144}{12}{12} has encoding rate $0.083$ and check weight $6$, against encoding rates of $0.055$ to $0.37$ and check weights of $10$ to $12$ for the evolved codes.

\textbf{Evolution under different mutation operators.} To assess the contribution of the LLM, we compare gpt-6.1-sol with two non-LLM mutations: random mutation and hill-climbing. Both retain the same evolution framework, seed programs, archive rules, evaluator and budget of $500$ steps. We perform five runs of each non-LLM mutation for each objective. Random mutation resamples the protograph entries at every step. At mutation levels 2 and 3, it additionally changes the protograph shapes or the group, respectively, using the same mutation-level selection rule as the LLM runs. At levels 2 and 3, hill-climbing changes the protograph shapes or the group in the same way as random mutation. At level 1, it applies one to three edits to the best measured code of the parent program, where an edit replaces, adds or removes one group element in a protograph entry (Supplemental Material~S2). We also examine how the search performance depends on the LLM and its reasoning effort. We replace gpt-6.1-sol with gpt-6-luna at low, high and extra-high reasoning effort, or with gpt-6-sol at high reasoning effort. These LLM comparisons use the same framework, seed programs, archive rules, evaluator and $500$-step budget, with one run per configuration for each objective.

\begin{figure}[t]
  \centering
  \includegraphics[width=\columnwidth]{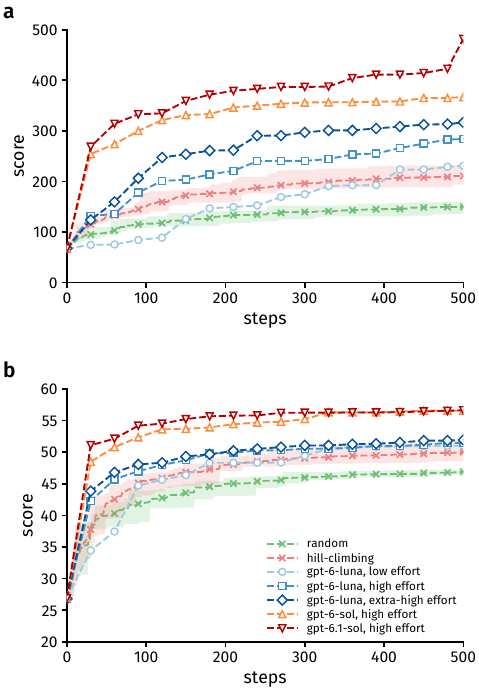}
  \caption{\textbf{Run score against steps for different mutation operators.} Sum over the ten encoding-rate cells of the best program score so far, for the FOM objective (\textbf{a}; $\mathcal{S}_{\mathrm{FOM}}$, Eq.~\eqref{eq:score}) and for the CLER objective (\textbf{b}; $\mathcal{S}_{\mathrm{CLER}}$, Eq.~\eqref{eq:plscore}). Each step is one model call or one random proposal. All runs start from the same eight seed programs. Markers are drawn every $30$ steps and at step $500$, joined by dashed lines. The random-mutation and hill-climbing curves are the mean over five runs, with the band from the minimum to the maximum, and the curve of every other source of mutations is one run. In (b), the gpt-6-sol and gpt-6.1-sol curves differ by less than $0.3$ from step $341$ onwards and overlap in the plot.
}
  \label{fig:trajectories}
\end{figure}

Fig.~\ref{fig:trajectories} shows the evolution of the run scores. Hill-climbing improves on random mutation for both objectives, indicating the benefit of local refinement over random-mutation. All LLM runs finish above all random-mutation runs, and the gpt-6-sol and gpt-6.1-sol runs also exceed every hill-climbing run on both objectives. The comparison with hill-climbing is more mixed for gpt-6-luna: its high- and extra-high-effort runs outperform all hill-climbing runs on the FOM objective, whereas its CLER scores remain close to the upper end of the hill-climbing range. Among the LLM configurations, increasing the reasoning effort of gpt-6-luna improves the final score on both objectives and reduces the number of steps needed to surpass the final mean score of the random-mutation. The gpt-6-sol and gpt-6.1-sol runs exceed this baseline earlier in the search than the gpt-6-luna runs. Their final performance depends on the objective: gpt-6.1-sol achieves the highest FOM score, while gpt-6-sol and gpt-6.1-sol reach nearly identical CLER scores.

\let\SCEFloatBarrier\FloatBarrier\let\FloatBarrier\relax

\section{Discussion}

SCE evolves both the algebraic setting and the internal search strategy of construction programs. A fixed external evaluator assesses multiple distinct candidate codes produced by each program. This search yields the exactly certified non-abelian \code{585}{121}{24} code at check weight $12$ and, after further optimization, \code{585}{117}{20} at check weight $9$. The FOM and circuit-level objectives favour different constructions, whose representative programs also use different internal search strategies. In the simulated memory experiment, a smaller-distance code can outperform a larger-distance code even at equal block length, encoding rate, check weight, CNOT count and circuit depth. More broadly, we expect that SCE could yield reusable construction strategies that can be adapted to different algebraic settings and design objectives.

Several directions follow. First, the circuit-level objective can incorporate the connectivity constraints of a target platform and the noise model of a specific device, steering the search toward hardware-compatible codes. Co-designing a code together with its syndrome-extraction circuit and decoder would extend SCE to a joint search over all three components. Second, the program-evolution framework could be adapted to other code families, such as quantum Tanner codes, fibre-bundle codes or pair-partition codes, by modifying the construction rules, mutation instructions and validity checks. Whether the search can discover entirely new CSS constructions remains an open question.

\section{Methods}

\textbf{Prompt.}
The search operates on construction programs rather than on individual codes.
A program fixes a group, the protograph shapes and a rule that assigns group-algebra entries to $A$ and $B$, and it generates candidate codes when executed.
In each step, one program is drawn from the archive as the parent, and the LLM modifies it to produce a child program.
Each call to the LLM consists of a system message and a user message.

The system message is identical in every step of a run.
It states the task and the construction grammar, for example the program format, the relation $n=fq$ between the number of physical qubits, the group order and the protograph shapes, and which protograph shapes are admissible.
It also defines the ten encoding-rate cells of the archive (described below), the score and the constraints on admissible codes.

The user message is assembled in every step from the records of the run and contains four parts.
The first is the parent program and its evaluation, which reports $k$, the check weight, the distance estimates and the score of every candidate and, for the measured candidates, the lightest logical operators found.
For the circuit-level objective, it also reports the circuit depth, the CNOT count and the logical error rate.
The second is the search status, which lists every pair of group and protograph shapes tried so far with its best score, and the best program of each encoding-rate cell.
The third is the lineage of the parent (up to three generations) and up to nine archive programs with their evaluations, namely the five best-scoring programs of the parent's cell and the best programs of four other cells, drawn uniformly at random.
The fourth is a mutation-level instruction that names the part of the program the LLM may change in this step: the rule that assigns the entries (level 1), the protograph shapes (level 2), or the group (level 3). In ordinary steps, the level is drawn with probability 1/3 each, and only the level-2 instruction names a target rate cell (Supplemental Material S3).

We use a MAP-Elites quality-diversity archive~\cite{mouret2015illuminating} with ten encoding-rate cells, numbered $0$ to $9$ and centred at $k/n=0.10,0.15,\ldots,0.55$.
Each cell maintains a pool of its $12$ best programs, and a program is assigned to the cell that corresponds to the mean encoding rate of its top $M$ measured codes.
In most steps, the parent is sampled from a target cell, which is scheduled by repeatedly shuffling the ten cells and visiting each once per cycle, so that every cell receives roughly the same number of steps over complete cycles.
If the target cell is empty, the parent is sampled from the nearest occupied cell.
Evaluated programs enter the archive, and the run score is the sum over the ten cells of the best program score.

\textbf{In-loop distance estimation.}
The evolution framework requires a fast and accurate distance estimator. The evaluator uses this estimator for candidate screening under both objectives and for program scoring under the FOM objective. It returns upper bounds on the code distance. Exact certification performed after evolution is described in Supplemental Material~S5. We develop a GPU implementation of a code-distance estimator based on the syndrome-decoder method of refs.~\cite{bravyi2024,webster2026}. The idea is to use a decoder to find a nontrivial logical operator
that acts on as few physical qubits as possible. The smallest
verified weight found gives an upper bound on the code distance. We describe the search for the $Z$-distance $d_Z$ as an example.
All matrix and vector operations are over $\mathbb{F}_2$.
Let $H=H_X$ and $S=H_Z$ be the matrices of $X$-type and
$Z$-type stabilizer generators, respectively, and let the rows
$L_1,\dots,L_k$ of $L=L_X$ represent a complete basis of
logical $X$ operators. A binary column vector $v$ represents a $Z$-type Pauli operator,
whose weight $\operatorname{wt}(v)$ is the number of physical qubits on which it acts nontrivially. The condition $Hv=0$
ensures that this operator commutes with every $X$-type
stabilizer. Commutation with all $Z$-type stabilizers is automatic.
However, this condition also admits the zero vector and
stabilizers. To exclude these trivial solutions, we additionally
require $L_i v=1$. This forces the candidate to anticommute
with the $i$th logical $X$ operator, so it must be a nontrivial
logical $Z$ operator.

For each $i=1,\dots,k$, we therefore search for low-weight
solutions of
\begin{equation}
\label{k-equations}
    \begin{bmatrix}
        H \\
        L_i
    \end{bmatrix}v
    =
    (0,\dots,0,1)^{\mathsf T}.
\end{equation}
Because the $L_i$ form a complete logical $X$ basis,
every nontrivial logical $Z$ operator satisfies $L_i v=1$
for at least one $i$. Consequently,
\begin{equation}
    d_Z =
    \min_{1\leq i\leq k}
    \ \min_{\substack{
        v\in\mathbb{F}_2^n \\
        Hv=0,\; L_i v=1
    }}
    \operatorname{wt}(v).
\end{equation}

To find low-weight solutions of these $k$ linear systems,
we use normalized min-sum belief propagation~\cite{Chen2002Near}. In the first phase, we solve all $k$ equations of Eq.~\ref{k-equations} once. At each decoding trial, belief propagation assumes the same bit-error $p$ on every qubit. Under this assumption, a small $p$ favours low-weight error patterns. Here we take the error rate  $p=0.05$. We run normalized min-sum belief propagation with a normalization factor of $0.8$ on the check-node messages for $24$ iterations to obtain posterior log-likelihood ratios of each qubit. Ordered-statistics decoding (OSD)~\cite{Panteleev2021Degenerate} then sorts the qubits by their log-likelihood ratios, from the most to the least likely to be in error. In this order, it selects each qubit whose column is linearly independent of the columns already selected, until the selected columns are enough to
satisfy the linear system. It sets all unselected qubits to zero and solves for the selected ones, which gives one
solution supported on the qubits most likely to be in error. Because belief propagation may underestimate the error probability of some unselected qubits, we also take the $48$ unselected qubits that are most
likely to be in error according to their log-likelihood ratios. For each of these $48$ qubits, we set its entry of $v$ to $1$, keep the entries of all other unselected qubits at zero, and solve the linear system again. We repeat this for all the $48$ qubits. Then, we do the same for every two-qubit pair of these $48$ qubits. Together with the first solution, this gives $1+48+1128=1177$ solutions of the linear
system, and we keep the lightest one~\cite{Roffe2020Decoding}. We then lower its weight further by a greedy local descent, where we add any single row or pair
of rows of $S$ that lowers the weight of $v$, and repeat this for at most six rounds.  These rows are stabilizers, so $v$ remains a nontrivial logical operator. We repeat such trials for all $k$ equations of Eq.~\ref{k-equations} at the same error rate. Each time we solve a batch of $128$ equations on the GPU.

The first phase ends after half of the time available for this distance or all $k$ equations have been solved. In the second phase, we sample each batch of $128$ equations with replacement from the $k$ equations randomly, and decode them again with $p$ drawn log-uniformly in $[0.003,0.35]$ independently for each qubit. We set the sampling with a preference for those that have yielded lightest solutions so far, where $70\%$ of equations are sampled from the $\max(8,\lfloor k/6\rfloor)$ equations with the lightest solutions and $30\%$ of sampling is uniformly drawn from all $k$ equations. A different $p$ makes the decoder return a different solution, so the repeated trials can
find lighter logical operators.

Every returned operator is checked again on the CPU ($Hv=0$ and $Lv\neq0$), so each reported weight is attained by an explicit logical operator and is an upper bound on the distance. The $X$-distance is obtained in the same way, in the remaining half of the time budget, with $H_X$ and $H_Z$ exchanged and $L_X$ replaced by $L_Z$, and the code distance is $d=\min(d_X,d_Z)$. For the reported codes with certified distances, the in-loop estimate equals the certified distance in most cases.

\textbf{Circuit-level protocol.}
We evaluate each code through a circuit-level quantum memory simulation with six rounds of syndrome extraction, using one ancilla qubit per stabilizer check. For the lifted-product codes considered here, each entry of a protograph expands into a $q\times q$ block, where $q=|G|$ is the group order. Each nonzero block connects a check block to a data block and is a sum of permutation matrices corresponding to distinct elements of $G$. We partition the nonzero entries of each block into sets, each containing exactly one entry in every row and every column. Each set is a perfect matching and defines a layer of $q$ CNOT gates that share no qubits and can therefore be executed simultaneously. We assign each layer to a time step, allowing multiple layers to run simultaneously only if they share no qubits. The execution order of the layers must also satisfy the $X$--$Z$ commutation condition in ref.~\cite{zhang2026asc}, which requires that, for each pair of an $X$ check and a $Z$ check, the number of shared data qubits on which the $X$ interaction precedes the $Z$ interaction be even. This condition ensures that the circuit measures the intended stabilizers. A qubit takes part in at most one CNOT gate per time step, so the depth is at least the largest number of CNOT gates acting on a single qubit. We use the constraint-programming solver CP-SAT of OR-Tools~\cite{perron2024cpsat}, which tests the depths in increasing order from this lower bound and returns the first depth for which all conditions can be met. Without noise, every detector of a correct circuit should give the same outcome in every shot. We use Stim~\cite{gidney2021stim} to simulate the circuits and check that the circuits are valid and measure the intended stabilizers.

The shots under depolarizing noise are decoded by a three-stage cascade. Each later stage runs only on the shots on which the previous stage fails to converge, so the slower stages handle only the few difficult shots, and the decoding remains fast enough for the circuit-level evaluation inside the search loop. The first two stages are the same decoder strategy as ref.~\cite{lu2026quantum}, which consists of a primary relay-BP decoder~\cite{muller2025relaybp}, and a secondary relay-BP when the first fails to converge. The third stage is belief propagation with localized-statistics decoding (LSD) post-processing~\cite{hillmann2024lsd} from the \texttt{ldpc} package~\cite{Roffe_LDPC_Python_tools_2022}. It runs on the CPU with at most $300$ min-sum belief-propagation iterations before LSD. At this stage, we use LSD instead of the ordered-statistics post-processing of ref.~\cite{lu2026quantum}, which would significantly slow down the in-loop evaluation. Because different decoders and schedules give different logical error rates, all evolved codes are simulated with the same schedule search and the same decoder, and the reference code \code{144}{12}{12} uses the published schedule of ref.~\cite{bravyi2024}.

For the sweep of Fig.~\ref{fig:sweep}, at each $p$ we sample both the $X$ and the $Z$ basis until $1000$ failures are recorded or a shot budget set for that $p$, between $5\times10^{4}$ shots at $p=0.005$ and $5\times10^{6}$ shots at $p=0.001$, is exhausted. For the codes with the lowest logical error rates at $p\le0.003$, the budget yields too few failures, and we sample further with up to $3\times10^{8}$ shots.

\textbf{Budgets and statistics.}
\begin{table*}[t]
\centering
\footnotesize
\setlength{\tabcolsep}{3.5pt}
\caption{\textbf{Running budget.} Call: median latency of an LLM request. Program: median execution time of a program on a single CPU core. Prompt/cached: median numbers of total and cached prompt tokens per model call, respectively. Response/reasoning: median numbers of total response and reasoning tokens per model call, respectively. CPU: single-core program execution time summed over the evaluated children. GPU: evaluation time summed over the evaluated children, with one GPU per evaluation. Each run evaluates up to $16$ children concurrently, each on a separate GPU, and allows up to $16$ concurrent LLM requests. Each evaluation uses $8$ CPU cores for the FOM objective and $16$ for the CLER objective. Computations use Intel Xeon Gold 6130 and Xeon Gold 6144 CPUs and NVIDIA Quadro RTX 6000 GPUs with 24 GB of memory. Cut-off: the knowledge cut-off stated in the model documentation. We note web search was disabled for every LLM call. Time: total search time of the run. For random mutation and hill-climbing, each reported value is the median over five runs.}
\label{tab:models}
\begin{tabular}{llrrrrrrl}
\toprule
Model (effort) & Objective & Call & Program & Prompt/cached & Response/reasoning & CPU / GPU & Time & Cut-off \\
 & & (s) & (s) & (tokens) & (tokens) & (h) & (h) & \\
\midrule
gpt-6.1-sol (high) & FOM  & 260 & 279 & 164k / 2.7k & 13.7k / 7.9k  & 59 / 45  & 15.7 & 2026-04-30 \\
                   & CLER & 308 & 301 & 156k / 3.0k & 14.3k / 8.3k  & 69 / 480 & 46.0 & \\
gpt-6-sol (high)   & FOM  & 166 & 25  & 120k / 2.7k & 10.2k / 6.7k  & 7 / 48   & 6.3 & 2026-04-20 \\
                   & CLER & 146 & 12  & 124k / 3.0k & 9.6k / 6.2k   & 2 / 597  & 53.5 & \\
gpt-6-luna (low)   & FOM  & 13  & 1   & 101k / 2.7k & 1.5k / 0.3k   & 0.4 / 39 & 4.2 & 2026-05-18 \\
                   & CLER & 13  & $<1$ & 117k / 3.0k & 1.5k / 0.4k  & 0.1 / 264 & 38.3 & \\
gpt-6-luna (high)  & FOM  & 56  & 32  & 114k / 2.7k & 6.8k / 4.9k   & 12 / 40  & 10.2 & \\
                   & CLER & 60  & 24  & 121k / 3.0k & 7.4k / 5.4k   & 6 / 256  & 36.6 & \\
gpt-6-luna (extra-high) & FOM & 108 & 59 & 124k / 2.7k & 12.8k / 10.4k & 13 / 42 & 11.4 & \\
                   & CLER & 102 & 30  & 125k / 3.0k & 13.1k / 10.9k & 9 / 259  & 36.9 & \\
Random mutation    & FOM  & --  & $<1$ & -- & -- & $<0.1$ / 45  & 3.6 & -- \\
                   & CLER & --  & $<1$ & -- & -- & $<0.1$ / 253 & 26.7 & \\
Hill-climbing      & FOM  & --  & $<1$ & -- & -- & $<0.1$ / 46  & 3.5 & -- \\
                   & CLER & --  & $<1$ & -- & -- & $<0.1$ / 328 & 24.7 & \\
\bottomrule
\end{tabular}
\end{table*}
The cost of the LLM runs is around \$270 and \$230 for the two gpt-6.1-sol runs, \$180 and \$170 for the two gpt-6-sol runs, and \$5--\$10
for each gpt-6-luna run.  For the FOM objective, the median evaluation time of a child is $305$--$315$\,s on one GPU in every run. Summed over the children, the evaluation takes $39$--$48$ GPU-hours for a language-model run and $44$--$45$ GPU-hours for a random-mutation run (Table~\ref{tab:models}).  The five random-mutation runs together used $223$ GPU-hours, five times the $45$ GPU-hours of the gpt-6.1-sol run.   The programs generated by gpt-6.1-sol are more complex, and their execution time is comparable to their evaluation time. For random mutation, in contrast, the runtime is dominated by the evaluation. Consequently, the search time of the gpt-6.1-sol run, $15.7$\,h, is about four times the median search time of a random-mutation run, $3.6$\,h (Table~\ref{tab:models}), and is shorter than the total time of the five random-mutation runs. On both objectives, the programs generated by gpt-6-sol run for a shorter time than those generated by gpt-6-luna at high and extra-high effort (Table~\ref{tab:models}). For the FOM objective, the median running time of a gpt-6-sol program is $25$\,s, against $32$\,s and $59$\,s for gpt-6-luna. For the CLER objective, these times are $12$\,s, $24$\,s and $30$\,s. The higher scores of gpt-6-sol therefore cannot be attributed to longer-running programs alone.

\section*{Data availability.} The data for this work is available in the repository~\cite{repo}.

\section*{Code availability.} The code for this work is available in the repository~\cite{repo}.

\section*{Author contributions.} F.M. proposed and supervised the project. Z.L. carried out the research. Both authors wrote the manuscript. 

\section*{Competing interests.} The authors declare no competing interests.

\section*{Acknowledgements.} Z.L. thanks Xie-Hang Yu for fruitful discussion. This research is part of the Munich Quantum
Valley, which is supported by the Bavarian state government with funds from
the Hightech Agenda Bayern Plus.

\section*{Use of AI tools}
Claude Code (Fable 5/5.1 and Opus 5/5.1/5.5) and Codex (GPT-6.1-sol, GPT-6-astra) assisted in developing the code of SCE, and large language models were used to polish the text. The authors reviewed all code and text and take full responsibility for the content of this work.

\clearpage
\onecolumngrid
\setcounter{figure}{0}
\setcounter{table}{0}
\setcounter{equation}{0}
\renewcommand{\thefigure}{S\arabic{figure}}
\renewcommand{\thetable}{S\arabic{table}}
\renewcommand{\theequation}{S\arabic{equation}}
\renewcommand{\theHfigure}{S\arabic{figure}}
\renewcommand{\theHtable}{S\arabic{table}}
\renewcommand{\theHequation}{S\arabic{equation}}
\setcounter{section}{0}
\renewcommand{\thesection}{S\arabic{section}}
\renewcommand{\theHsection}{S\arabic{section}}
\setcounter{secnumdepth}{1}

\begin{center}{\large \textbf{Supplemental Material}}\end{center}

\section{Lifted-product construction and CSS validity}\label{supp:lp}
\paragraph*{Calderbank--Shor--Steane (CSS) codes.}
A CSS code~\cite{Calderbank1996Good,Steane1996Error} is specified by two
binary parity-check matrices $H_X$ and $H_Z$ with the same number of
columns (number of physical qubits).  The rows of $H_X$ define $X$-type stabilizers, and the rows of
$H_Z$ define $Z$-type stabilizers.  The two sets of stabilizers commute
if and only if
\begin{equation}
H_X H_Z^{\mathsf T}=0 \pmod 2 .
\label{eq:css}
\end{equation}
If the matrices have $n$ columns, then the code has block length $n$ and
encodes
\[
k=n-\operatorname{rank}_2 H_X-\operatorname{rank}_2 H_Z
\]
logical qubits.  Thus, a central task in constructing CSS codes is to
generate parity-check matrices for which Eq.~\eqref{eq:css} holds.  

\paragraph*{Hypergraph product (HGP) codes.}
 Let
\[
A\in\Ftwo^{m_A\times n_A},
\qquad
B\in\Ftwo^{m_B\times n_B}
\]
be two classical binary parity-check matrices.  The hypergraph product
defines~\cite{tillich2014}
\begin{align}
H_X
&=
\bigl[\,A\otimes I_{n_B}
\;\big|\;
I_{m_A}\otimes B^{\mathsf T}\,\bigr],
\nonumber\\
H_Z
&=
\bigl[\,I_{n_A}\otimes B
\;\big|\;
A^{\mathsf T}\otimes I_{m_B}\,\bigr].
\label{eq:hgp}
\end{align}
Then $H_XH_Z^{\mathsf T}$ contains two identical contributions, $A\otimes B^{\mathsf T}+A\otimes B^{\mathsf T}$,
which cancel over $\Ftwo$.  Hence $H_XH_Z^{\mathsf T}=0$.  In other
words, the CSS condition follows from the two-sector structure of
Eq.~\eqref{eq:hgp}. 

\paragraph*{Lifted-product codes.}
 Lifted-product codes preserve the same mechanism as HGP: the product
$H_XH_Z^{\mathsf T}$ again decomposes into two matching contributions
that cancel over $\mathbb F_2$, while the scalar entries of $A$ and $B$
are replaced by structured permutation blocks.

Let
\[
A\in \mathbb F_2[G]^{m_A\times n_A},\qquad
B\in \mathbb F_2[G]^{m_B\times n_B}
\]
be two small matrices over the group algebra of a finite group $G$ of order $q = |G|$. An entry of $A$ (and likewise $B$) is a formal sparse sum $a=\sum_{g\in G} a_g g$ with the coefficient $a_g\in\mathbb F_2$. To
obtain ordinary binary matrices, each such group element is
replaced by a $q\times q$ permutation block sum. More explicitly, we use
the left regular representation for entries of $A$ and the right regular
representation for entries of $B$,
\[
a\mapsto \lambda(a)=\sum_{g\in G} a_g\,\lambda(g),\qquad
b\mapsto \rho(b)=\sum_{g\in G} b_g\,\rho(g),
\]
where $\lambda(g)$ and $\rho(g)$ are the permutation matrices acting on
the $q$-dimensional space with basis $\{e_h\}_{h\in G}$ by
\[
\lambda(g)e_h=e_{gh}, \qquad \rho(g)e_h=e_{hg^{-1}}
\qquad (g,h\in G).
\]
A direct check gives
\begin{equation}
[\lambda(a),\rho(b)]=0
\qquad\text{for all } a,b\in \mathbb F_2[G],
\label{eq}
\end{equation}
which is the property responsible for the cancellation in
$H_XH_Z^{\mathsf T}$.

To build the code, we replace each matrix entry by its regular representation block:
\[
A_{ij}\ \longmapsto\ \lambda(A_{ij}),
\qquad
B_{ij}\ \longmapsto\ \rho(B_{ij}).
\]
An $m_A\times n_A$ array of group-algebra entries thereby becomes an
$m_A\times n_A$ array of sparse $q\times q$ binary blocks, i.e.\ a matrix
$\widetilde A\in\mathbb F_2^{\,m_Aq\times n_Aq}$. Likewise $B$ becomes
$\widetilde B\in\mathbb F_2^{\,m_Bq\times n_Bq}$. The CSS parity-check
matrices then follow the hypergraph-product rule applied to the block
(pre-lift) indices,
\begin{equation}
H_X=\bigl[\,\widetilde A\otimes I_{n_B}\ \big|\ I_{m_A}\otimes \widetilde B^{\mathsf T}\,\bigr],
\qquad
H_Z=\bigl[\,I_{n_A}\otimes \widetilde B\ \big|\ \widetilde A^{\mathsf T}\otimes I_{m_B}\,\bigr],
\label{eq:lp-sm}
\end{equation}
where each $\otimes I$ replicates a block index only. The two sectors yield a quantum code on $n=fq$
physical qubits, where $f=n_An_B+m_Am_B$.  By the
left--right commutation Eq.~\eqref{eq}, we have $H_XH_Z^{\mathsf T}=0$, so that
Eq.~\eqref{eq:lp-sm} defines a valid CSS code.

The construction above is the algebraic search space used in this work. A candidate produced by the search is specified compactly by a finite group $G$, two small sparse group-algebra matrices $A$ and $B$, and their protograph shapes. The lifting map then deterministically converts this symbolic description into binary CSS check matrices $H_X$ and $H_Z$.

\section{Seed programs and random mutation}\label{supp:prompts}
All runs reported in this paper start from the same eight seed programs, listed in Table~\ref{tab:seeds}.  Each seed program generates its candidate codes at random: it fills each entry with one to five distinct group elements drawn uniformly at random and removes terms at random until the check-weight cap is met.  Cells 0, 7, 8 and 9 hold no seed.  Under the FOM objective, the best seed scores of the cells sum to $67.6$.  Under the CLER objective, the seeds score $3.2$--$5.9$ each.  Random mutation keeps this program text unchanged and changes only the group, the protograph shapes, and the random seed of the entries.  Hill-climbing uses the same proposals as random mutation at the group and protograph shape mutation levels.  At the first level it starts from the best measured code of the parent and applies one, two or three edits to its protograph entries, where an edit replaces, adds or removes one group element in one entry.  Each of the sixty candidates is the parent code after its own edits, and edits that are equivalent to multiplying protograph rows or columns by group elements are drawn again.

\begin{table}[!htbp]
\centering
\caption{Seed programs.  The columns give the target cell, the group with its order in bracket, the protograph shapes, the block length and the FOM program score.}
\label{tab:seeds}
\begin{tabular}{llllr}
\toprule
Cell ($k/n$) & Group & $A\times B$ & $n$ & $\mathcal{S}_{\mathrm{FOM}}$ \\
\midrule
1 (0.15) & $C_{23}$ (23) & $(2,4)\times(3,5)$ & 598 & 6.7 \\
2 (0.20) & $D_{14}$ (28) & $(2,4)\times(2,4)$ & 560 & 15.9 \\
2 (0.20) & $D_{59}$ (118) & $(1,2)\times(1,2)$ & 590 & 20.4 \\
3 (0.25) & $C_{12}\rtimes C_2$ (24) & $(5,2)\times(4,2)$ & 576 & 14.7 \\
4 (0.30) & $C_{21}$ (21) & $(2,4)\times(2,6)$ & 588 & 10.5 \\
5 (0.35) & $C_{17}$ (17) & $(2,5)\times(2,6)$ & 578 & 8.8 \\
6 (0.40) & $D_{7}$ (14) & $(6,2)\times(6,2)$ & 560 & 6.5 \\
6 (0.40) & $D_{29}$ (58) & $(1,3)\times(1,3)$ & 580 & 6.4 \\
\bottomrule
\end{tabular}
\end{table}

\section{Novelty check, archive and parent selection}\label{supp:engine}
This section describes four parts of the evolution framework that the main text summarizes.  They are the
test of whether a candidate code is new, the organization of the MAP-Elites archive, islands, and the choice of
the parent of a step.  The evolution framework is our modification of OpenEvolve~\cite{openevolve}. 

\paragraph*{Novelty check.}
The evaluator receives the sixty candidate entry assignments of a program.  It first checks that the multiplication table of the group has an identity and inverses (the program itself checks associativity on sampled triples), and it then builds the pair $(H_X,H_Z)$ of every
candidate from this table.  It discards candidates with $k=0$, with a check weight above the cap or with
$H_XH_Z^{\mathsf T}\neq0$.  A candidate that matches a code already in the archive, exactly or up to a relabelling of qubits and checks,
receives a score of zero.

\paragraph*{MAP-Elites archive.}
The archive uses the encoding rate $k/n$ as a feature dimension. The feature value of a program is the mean
$k/n$ of its five top-scoring codes, and the program is assigned to the nearest of ten centres $0.10,0.15,\ldots,0.55$. We note the first cell $k/n<0.125$ and the last cell $k/n>0.525$ are open-ended.  Each cell keeps its elite, which is the program with the highest
program score, and a pool of at most twelve programs ranked by program score.  A new program enters the
pool of its cell when it ranks among the top twelve, and it then replaces the lowest-ranked program.  The run score is the sum over the ten cells of the best program score recorded with the
evaluations.

\paragraph*{Island model and migration.}
The population is divided into $N_{\mathrm{isl}}=3$ islands. At the start of each run, the seed programs are distributed over the islands in turn (seed $i$ to island $i \bmod N_{\mathrm{isl}}$), and each child is placed on the island of its rate cell (cell $c$ to island $c \bmod N_{\mathrm{isl}}$). Migration is triggered when the largest island generation counter has advanced by $25$ since the previous migration event. From each island the top fraction $\mu=0.1$ by program score, and at least one program, is copied to its two ring-adjacent neighbours (islands $i\!\pm\!1 \bmod N_{\mathrm{isl}}$). To avoid duplication, a migrated copy is not migrated again.

\paragraph*{Parent of a step.}
Each step is assigned a target cell. The ten cells are taken in a random order that is drawn again every ten steps, so that every cell is targeted equally often. With probability $0.7$, the parent is drawn from the pool of the target cell, or from the pool of the nearest occupied cell if the target cell is empty, with probability proportional to $\exp(s/T)$, where $s$ is the program score and $T = 0.2$.
With probability $0.2$, the parent is drawn uniformly at random from the programs of one island, and with probability $0.1$ from all programs. When a level-2 or level-3 child becomes the best program of a cell other than its parent's, it is the parent of the next two steps. Ordinary steps draw the mutation level uniformly from levels 1, 2 and 3. Only the level-2 instruction names the target cell.

\section{Run with weight $\le9$ and $n<600$}\label{supp:w9}
We repeat the FOM run of gpt-6.1-sol with the check-weight cap lowered from 12 to 9. Every other setting remains the same.
Table~\ref{tab:w9cells} lists the best program of every cell with its best measured code.

In the rate-$0.2$ cell, the 258th child program constructs codes over $C_{13}\rtimes C_9$ with protograph shapes $(1,2)\times(1,2)$ and check weight $9$. It generates a \code{585}{119}{18} code whose distance has been certified exactly, giving a FOM of $kd^2/n=65.9$. This is comparable to the FOM of $64.8$ for the published weight-$9$ lifted-product code \code{540}{108}{18} at rate $0.2$~\cite{mitten2026}. A later program produces five top-ranked candidates, three codes \code{550}{110}{18}, one \code{550}{110}{17} and one \code{550}{110}{13}. Although its best FOM of $64.8$ is lower than $65.9$, the consistently strong scores across all five candidates keep the program among the elites in this cell.

\begin{table}[!htbp]
\centering
\footnotesize
\caption{Best code of every rate cell of the gpt-6.1-sol run with check weight at most 9, after 500 children.  Each row gives the code with the largest $kd^2/n$ among the five measured codes of the best program of the cell, its group, the protograph shapes and the check weight $w$. In the rate-$0.2$, the first row is the best code of the $258$-th child program. All distances are certified exactly (Supplemental Material~\ref{supp:exactd}).  The code marked R is generated by the fifth child of the refinement of the 258th child program with level-1 instructions only.}
\label{tab:w9cells}
\setlength{\tabcolsep}{4pt}
\begin{tabular}{llllc}
\toprule
Cell ($k/n$) & Group & $A\times B$ & Code & $w$ \\
\midrule
0 ($<0.125$) & $\mathrm{Dic}_{11}$ & $(2,3)\times(2,3)$ & \code{572}{48}{24} & 9 \\
1 (0.15) & $C_{31}$ & $(2,3)\times(2,4)$ & \code{496}{72}{14} & 9 \\
2 (0.20) & $C_{13}\rtimes C_{9}$ & $(1,2)\times(1,2)$ & \code{585}{119}{18} & 9 \\
 & $C_{13}\rtimes C_{9}$ & $(1,2)\times(1,2)$ & \code{585}{117}{20}$^{\mathrm{R}}$ & 9 \\
3 (0.25) & $C_3\rtimes C_{16}$ & $(1,2)\times(2,5)$ & \code{576}{150}{9} & 9 \\
4 (0.30) & $C_5\rtimes C_4$ & $(2,5)\times(2,5)$ & \code{580}{180}{7} & 9 \\
5 (0.35) & $C_8$ & $(3,8)\times(3,8)$ & \code{584}{200}{6} & 9 \\
6 (0.40) & $C_{11}\rtimes C_5$ & $(1,3)\times(1,3)$ & \code{550}{226}{5} & 8 \\
7 (0.45) & $C_{12}\rtimes C_2$ & $(1,3)\times(2,7)$ & \code{552}{258}{4} & 9 \\
8 (0.50) & $C_{11}$ & $(2,7)\times(2,7)$ & \code{583}{287}{4} & 9 \\
9 ($>0.525$) & $\mathrm{Dic}_7$ & $(1,4)\times(1,4)$ & \code{476}{252}{3} & 9 \\
\bottomrule
\end{tabular}
\end{table}

\begin{figure}[!htbp]
\centering
\includegraphics[width=8.9cm]{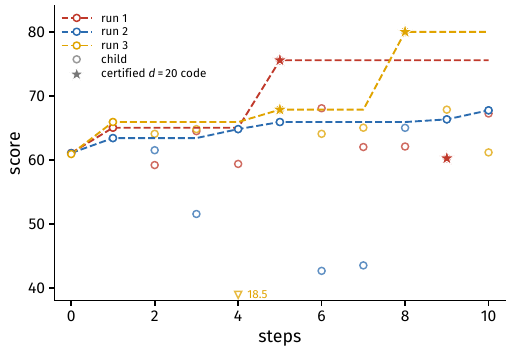}
\caption{In-loop program score $\mathcal{S}_{\mathrm{FOM}}$ over the first ten steps of three independent level-1 refinement runs that start from the 258th child of the weight-$9$ run. Each step is one model call, and the children are evaluated one at a time on one GPU. Dashed lines show the best score so far. Open circles show the score of each child. Stars mark children whose distance reaches $20$. Runs $1$ and $3$ find the \code{585}{117}{20} code at the fifth step, after $4.8$ and $4.1$ hours of wall-clock time, respectively. Run $2$ reaches its fifth step after $3.6$ hours.}
\label{fig:w9refine}
\end{figure}

We further refine the program of the 258th child in three independent runs. Every step receives only the level-1 instruction, and the target cell is fixed to the rate-$0.2$ cell, so each child changes only the rule that assigns the entries, while the group $C_{13}\rtimes C_9$ and the protograph shapes $(1,2)\times(1,2)$ stay those of the parent. Children are evaluated one at a time, and the code archive starts empty. All other settings are those of the run above. In two of the three runs, the fifth child generates codes \code{585}{117}{20} whose distances are certified exactly (Supplemental Material~\ref{supp:exactd}), two in one run and one in the other. Figure~\ref{fig:w9refine} shows the program scores of the first ten children of each run. The two fifth-child programs run for $56$ (Intel Xeon E5-2698 v4) and $91$\,min (Intel Xeon Gold 6130) on single CPU core. These codes reach $kd^2/n=80.0$ with $585$ qubits. Among the published weight-$9$ lifted-product codes at rate $0.2$~\cite{hong2026rate,mitten2026}, those with fewer than $600$ qubits have $d\le18$, and the shortest ones with a distance upper bound of $20$ are \code{625}{125}{$\le$20}~\cite{hong2026rate} and \code{630}{126}{$\le$20}~\cite{mitten2026}.

\section{Post-search exact-distance enumeration}\label{supp:exactd}

We certify minimum distances using the irreducible-cluster algorithm of ref.~\cite{dumer2017}, following the CSS procedure in Appendix~B.1 of ref.~\cite{lee2026pp}. The branching rule, the pruning bounds and the test of logical nontriviality follow that procedure. The main modification concerns the symmetry reduction used to select starting qubits: Appendix~B.1 uses cyclic translations of circulant lifts, whereas we use verified automorphisms of the Tanner graph, so the reduction does not require a cyclic lift. The search is performed separately for $X$- and $Z$-type logical operators. For each type, let $H$ be the opposite-type check matrix and $S$ the same-type stabilizer matrix. Thus, for $X$-type operators, $H=H_Z$ and $S=H_X$, with the roles reversed for $Z$-type operators. At a weight threshold $W$, the algorithm determines whether there exists a binary vector $x$ satisfying
\begin{equation}
    Hx=0,\qquad
x\notin\mathrm{rowspan}(S),\qquad
|x|\le W,
\end{equation}
where $|x|$ is the Hamming weight. The first condition ensures that the represented operator commutes with the stabilizers, and the second excludes stabilizer products, leaving a nontrivial logical operator.

The search builds a support qubit by qubit, up to weight $W$, while tracking its syndrome. When the syndrome is nonzero, it selects the unsatisfied check with the fewest qubits still available and tries adding each of these qubits in turn. After trying a qubit, it excludes that qubit from subsequent branches at the same step to avoid duplicate exploration. When the syndrome becomes zero, the support vector is reduced against a row-echelon basis of $S$. A nonzero remainder shows that the support represents a nontrivial logical operator of weight at most $W$. A zero remainder identifies a stabilizer, and the branch stops. Two lower bounds on the number of additional qubits prune the search. The first is the number of unsatisfied checks divided by the largest column weight of $H$, rounded up to the nearest integer. The second is obtained by greedily selecting unsatisfied checks whose sets of available qubits are pairwise disjoint. Each selected check requires a different additional qubit, so the number of selected checks is also a lower bound. A branch is pruned when its current weight plus the larger of these two bounds exceeds $W$. For a fixed logical type, a minimum-weight logical support has no nonempty proper subset with zero syndrome. Such a subset would either be a lighter logical operator or be a stabilizer whose removal leaves a lighter logical operator. Consequently, the support is connected in the Tanner graph and can be reached by the search starting from any of its qubits.

To reduce the number of starting qubits, we use nauty~\cite{nauty} to compute generators of the automorphism group of the Tanner graph that map qubits to qubits, $X$ checks to $X$ checks and $Z$ checks to $Z$ checks. We verify that the induced qubit permutations preserve the row spaces of both $H_X$ and $H_Z$. The group generated by these permutations partitions the qubits into orbits. We order these orbits and start the search from one representative in each, excluding all qubits in earlier orbits. This reduction preserves completeness. For any minimum-weight logical operator, consider the first orbit that its support meets. A permutation in the verified group maps a qubit of this support to that orbit's representative while preserving the operator's weight, logical type and nontriviality. Since each orbit is mapped onto itself, the transformed support still avoids all earlier orbits. It can therefore be reached by the search starting from that representative. If the search completes at weight level $W$ for both logical types without finding a nontrivial logical operator, it proves $d>W$. If a logical operator of weight $W+1$ has also been found and verified, the distance is certified as $d=W+1$.
\begin{table}[!htbp]
\centering
\footnotesize
\setlength{\tabcolsep}{5pt}
\caption{Reported codes and their verification status. Cell~0 of the FOM run lists the codes of its two two-block programs. The In-loop column reports the distance found during the run, using a search time limit of 30 seconds for FOM and 2 seconds for CLER. For the two cell-0 codes whose distances are upper bounds, the 600\,s and 2\,h columns report the results of two independent searches each, with time limits of 600 seconds and two hours. The Exact column reports the certified minimum distance. The dist-m4ri column reports the minimum distance computed independently using dist-m4ri~\cite{dumer2017}. In the Exact and dist-m4ri columns, a dash indicates that the computation did not finish within the allotted computational budget.}
\label{tab:hashes}
\resizebox{\ifdim\width>\linewidth\linewidth\else\width\fi}{!}{%
\begin{tabular}{cllllllll}
\toprule
Cell & Code & Group & Run & In-loop & 600\,s & 2\,h & Exact & dist-m4ri \\
\midrule
0 & $[[578,32]]$  & $C_{17}\times C_{17}$ & FOM & 46 & 46, 46 & 46, 46 & -- & -- \\
0 & $[[510,40]]$  & $C_{255}$             & FOM & 48 & 46, 46 & 42, 42 & -- & -- \\
1 & $[[592,83]]$  & $C_{37}$              & FOM & 18 & -- & -- & 18 & 18 \\
2 & $[[585,121]]$ & $C_{13}\rtimes C_9$   & FOM & 24 & -- & -- & 24 & -- \\
3 & $[[560,130]]$ & $\mathrm{Dic}_7$      & FOM & 14 & -- & -- & 14 & 14 \\
4 & $[[588,176]]$ & $\mathrm{Dic}_{21}$   & FOM & 12 & -- & -- & 12 & 12 \\
5 & $[[544,200]]$ & $Q_{16}$              & FOM & 8  & -- & -- & 8  & 8  \\
6 & $[[560,224]]$ & $\mathrm{AGL}(1,8)$   & FOM & 10 & -- & -- & 10 & 10 \\
7 & $[[483,220]]$ & $C_7\rtimes C_3$      & FOM & 6  & -- & -- & 6  & 6  \\
8 & $[[520,252]]$ & $C_5\rtimes C_8$      & FOM & 5  & -- & -- & 5  & 5  \\
9 & $[[504,297]]$ & $S_3\times C_4$       & FOM & 4  & -- & -- & 4  & 4  \\
\midrule
0 & $[[572,67]]$  & $\mathrm{Dic}_{11}$              & CLER & 22 & -- & -- & 22 & -- \\
1 & $[[576,82]]$  & $\mathrm{Dic}_9$                 & CLER & 15 & -- & -- & 15 & 15 \\
2 & $[[580,116]]$ & $C_{29}$                         & CLER & 12 & -- & -- & 12 & 12 \\
3 & $[[580,132]]$ & $C_{29}$                         & CLER & 14 & -- & -- & 14 & 14 \\
4 & $[[588,176]]$ & $C_2\times(C_7\rtimes C_3)$      & CLER & 10 & -- & -- & 10 & 10 \\
5 & $[[598,195]]$ & $C_{13}$                         & CLER & 7  & -- & -- & 7  & 7  \\
6 & $[[550,220]]$ & $C_{11}\rtimes C_5$              & CLER & 8  & -- & -- & 7  & 7  \\
7 & $[[552,240]]$ & $C_3\rtimes C_8$                 & CLER & 5  & -- & -- & 5  & 5  \\
8 & $[[570,288]]$ & $C_{19}$                         & CLER & 4  & -- & -- & 4  & 4  \\
9 & $[[544,288]]$ & $\mathrm{SD}_{32}$               & CLER & 4  & -- & -- & 4  & 4  \\
\midrule
0 & $[[572,48]]$  & $\mathrm{Dic}_{11}$   & FOM, $w\le9$     & 24 & -- & -- & 24 & -- \\
1 & $[[496,72]]$  & $C_{31}$              & FOM, $w\le9$     & 14 & -- & -- & 14 & 14 \\
2 & $[[585,119]]$ & $C_{13}\rtimes C_9$   & FOM, $w\le9$     & 18 & -- & -- & 18 & 18 \\
2 & $[[585,117]]$ & $C_{13}\rtimes C_9$   & FOM, $w\le9$ (R) & 20 & -- & -- & 20 & 20 \\
3 & $[[576,150]]$ & $C_3\rtimes C_{16}$   & FOM, $w\le9$     & 9  & -- & -- & 9  & 9  \\
4 & $[[580,180]]$ & $C_5\rtimes C_4$      & FOM, $w\le9$     & 7  & -- & -- & 7  & 7  \\
5 & $[[584,200]]$ & $C_8$                 & FOM, $w\le9$     & 6  & -- & -- & 6  & 6  \\
6 & $[[550,226]]$ & $C_{11}\rtimes C_5$   & FOM, $w\le9$     & 5  & -- & -- & 5  & 5  \\
7 & $[[552,258]]$ & $C_{12}\rtimes C_2$   & FOM, $w\le9$     & 4  & -- & -- & 4  & 4  \\
8 & $[[583,287]]$ & $C_{11}$              & FOM, $w\le9$     & 4  & -- & -- & 4  & 4  \\
9 & $[[476,252]]$ & $\mathrm{Dic}_7$      & FOM, $w\le9$     & 3  & -- & -- & 3  & 3  \\
\bottomrule
\end{tabular}}
\end{table}

\section{Representative program summaries}\label{supp:design}
\begin{figure}[!htbp]
\centering
\begin{tikzpicture}[
  font=\small,
  fbox/.style={draw=black!45, fill=white, rounded corners=2pt, align=flush left, text width=11.5cm, inner sep=5pt, line width=0.5pt},
  fscr/.style={fbox, fill=black!8},
  fout/.style={draw=black!55, fill=black!18, rounded corners=2pt, align=center, text width=11.5cm, inner sep=5pt, line width=0.5pt},
  farr/.style={-{Stealth[length=4pt]}, black!65, line width=0.6pt},
  node distance=3.2mm]
\node[fbox] (f1) {Prescribe common factors of $a(x)$, $b(x)$ to bound $k$};
\node[fbox, below=of f1] (f2) {List the polynomials; sample pairs with $32\le k<n/8$};
\node[fscr, below=of f2] (f3) {Stage 1: lightweight RIS search, 2 rounds; keep the best 8 by $kd^{2}/n$};
\node[fscr, below=of f3] (f4) {Stage 2: longer RIS search, 4 rounds + 4 $C_3$-restricted rounds; keep the best 4};
\node[fscr, below=of f4] (f5) {Stage 3: even longer RIS search, 16 rounds + 12 $C_3$-restricted rounds; keep the best};
\node[fout, below=of f5] (f6) {Candidate code};
\foreach \u/\v in {f1/f2,f2/f3,f3/f4,f4/f5,f5/f6} \draw[farr] (\u) -- (\v);
\end{tikzpicture}
\caption{Steps by which the FOM elite program of the rate cell $k/n<0.125$ generates one candidate code.  White boxes are the algebraic design and grey boxes the three stages of the distance estimator that the program implements itself.  RIS denotes a random information set search.  One round is one Gaussian elimination on randomly ordered columns, counted per candidate pair and per basis, and a $C_3$-restricted round considers only logical operators that are invariant under the subgroup $C_3$ of order $3$ generated by $x^{85}$.}
\label{fig:fomflow}
\end{figure}
This section expands the two programs discussed in the main text. One is the FOM elite program of the rate cell $k/n<0.125$, which builds generalized bicycle codes over the cyclic group $C_{255}$ with protograph shapes $(1,1)\times(1,1)$ and generates the code \code{510}{40}{$\le$42} of Table~1. The other is the CLER elite program of the same cell, which builds lifted-product codes over the dicyclic group $\mathrm{Dic}_{11}$ with protograph shapes $(2,3)\times(2,3)$ and generates the code \code{572}{67}{22} of Table~1. For each program, we show the steps by which it generates one candidate code, describe these steps and reproduce the program's design statement verbatim.

The workflow of the FOM program is shown in Fig.~\ref{fig:fomflow}. The FOM program builds generalized bicycle codes over the cyclic group $C_{255}$, which is generated by $x$ with $x^{255}=e$.  Both $A=a(x)$ and $B=b(x)$ are single polynomials with five terms, so that every check has weight $10$.  For these codes, $k=2\deg\gcd\bigl(a(x),b(x),x^{255}-1\bigr)$~\cite{lin2024quantum}.  The program therefore requires $a(x)$ and $b(x)$ to share three irreducible factors of $x^{255}-1$, of degrees $8$, $8$ and $4$ or $8$, $8$ and $2$, which guarantees $k\ge40$ or $k\ge36$.  It lists all five-term polynomials with these factors and removes those whose exponents all lie in a proper subgroup of $C_{255}$.  For each candidate, the program samples up to $56$ pairs $(a(x),b(x))$ from this list and keeps the pairs with $32\le k<n/8$.  It then estimates the distance $d$ of each pair with a three-stage estimator that the program implements itself.  The first stage uses a lightweight random search for logical operators and keeps the eight pairs with the largest $kd^{2}/n$.  At the second stage, the program gives these eight pairs a longer search.  Besides the ordinary search, a second search uses the symmetry of the group: it looks only for logical operators that are invariant under the subgroup of order $3$ generated by $x^{85}$, which reduces the search space.  At the third stage, the program keeps the best four pairs and gives them an even longer search of both kinds.  Finally, it ranks the four pairs again and returns the best one as the candidate, skipping any code that it has already returned for another candidate.  On a single core of a Xeon Gold 6144 (3.5 GHz), the program generates all sixty candidates in $328$\,s.

\begin{figure}[!htbp]
\centering
\begin{tikzpicture}[
  font=\small,
  fbox/.style={draw=black!45, fill=white, rounded corners=2pt, align=flush left, text width=11.5cm, inner sep=5pt, line width=0.5pt},
  fscr/.style={fbox, fill=black!8},
  fout/.style={draw=black!55, fill=black!18, rounded corners=2pt, align=center, text width=11.5cm, inner sep=5pt, line width=0.5pt},
  farr/.style={-{Stealth[length=4pt]}, black!65, line width=0.6pt},
  node distance=3.2mm]
\node[fbox] (c1) {Fill each entry of $A$ with $g+x^{t}g$ ($t$ odd, $t\neq11$) and of $B$ with $x^{a}+x^{a'}y$};
\node[fbox, below=of c1] (c2) {Sample 96 random fillings; keep those with $60\le k\le70$};
\node[fscr, below=of c2] (c3) {Reject if the classical code of $A$ or $B$ has a nonzero codeword of weight $\le4$};
\node[fscr, below=of c3] (c4) {Keep the 6 fillings with the fewest short cycles};
\node[fscr, below=of c4] (c5) {Rank by classical distances (RIS, 40 rounds, cap 22), then trapping-set score};
\node[fscr, below=of c5] (c6) {Check in rank order: no logical operator of weight $\le4$; keep the first pass};
\node[fout, below=of c6] (c7) {Candidate code};
\foreach \u/\v in {c1/c2,c2/c3,c3/c4,c4/c5,c5/c6,c6/c7} \draw[farr] (\u) -- (\v);
\end{tikzpicture}
\caption{Steps by which the CLER elite program of the rate cell $k/n<0.125$ generates one candidate code.  White boxes are the construction of the fillings and grey boxes the screening that the program implements itself.  The group $\mathrm{Dic}_{11}$ is generated by $x$ of order $22$ and $y$, with $y^{2}=x^{11}$ and $yxy^{-1}=x^{-1}$. $g$ is a group element, and the six values of $t$ in $A$ follow a pattern from a list that the program generates itself.  RIS denotes a random information set search, and one round is one Gaussian elimination on randomly ordered columns.}
\label{fig:clerflow}
\end{figure}

The workflow of the CLER program is shown in Fig.~\ref{fig:clerflow}. The program fills every entry of $A$ and $B$ with two elements of the dicyclic group $\mathrm{Dic}_{11}$, which is generated by $x$ of order $22$ and $y$, with $y^{2}=x^{11}$ and $yxy^{-1}=x^{-1}$.  In each entry of $A$, the second group element is $x^{t}$ times the first element, with $t$ odd and $t\neq11$.  The six values of $t$ follow one of the patterns in a list that the program generates itself.  In each entry of $B$, one element has the form $x^{a}$ and the other has the form $x^{a'}y$.  For each candidate, the program samples random fillings according to these rules and keeps those with $60\le k\le70$.  Then, the program rejects fillings in which the classical code defined by $A$ or by $B$ alone has a nonzero codeword of weight at most $4$.  Among the remaining fillings, it keeps the six with the fewest short cycles in the Tanner graphs of the two classical codes.  It then ranks them first by the distances of the two classical codes, estimated by a lightweight random search and capped at $22$.  Fillings that reach the cap are further ranked by a trapping-set score~\cite{raveendran2021trapping}.  Finally, the program checks the six fillings in this order and returns the first one whose quantum code has no logical operator of weight at most $4$ as the candidate.  On a single core of a Xeon Gold 6144 (3.5 GHz), the program generates all sixty candidates in $80$\,s.

Below we reproduce verbatim the structured specifications of these two programs. The fields \textsc{Concept}, \textsc{Group}, \textsc{Shapes} and \textsc{Filling} describe the construction idea, the finite group, the protograph shapes and the entry-assignment rule, respectively. These statements record the model's design rationale. The code parameters and performance reported in this manuscript are established by the external evaluation and verification procedures. The full programs are available in the repository~\cite{repo}.

We note some terms in these statements differ from the terminology used elsewhere in this manuscript. For clarity, we explain their meanings here. A \emph{bed} consists of a group together with a set of protograph shapes, and \texttt{BEDS} is the list of beds evaluated so far during a run. A \emph{slot} represents a candidate code. A \emph{card} specifies the mutation-level instruction for a step, with L3, L2, and L1 denoting levels~3, 2, and~1, respectively. In the FOM statement, 'twenty cards' denotes the twenty design configurations assigned to the candidates. The labels \texttt{d2s} and \texttt{d30s} denote distance estimates obtained with two-second and thirty-second time budgets, respectively. The \emph{fitness} is the program score. \texttt{cell 0.100} denotes the rate cell centred at $k/n=0.10$, and \texttt{kb} denotes the lower bound on $k$ guaranteed by the protograph shapes.

The qubits are divided into two sectors. Sector one contains $n_A n_B q$ qubits, arranged in blocks of $q$ qubits indexed by pairs consisting of a column of $A$ and a column of $B$. Sector two contains $m_A m_B q$ qubits, similarly arranged in blocks indexed by pairs consisting of a row of $A$ and a row of $B$. For example, \texttt{Acol0/Bcol0} labels the block in sector one associated with column~0 of $A$ and column~0 of $B$, while \texttt{Arow0/Brow0} labels the block in sector two associated with row~0 of $A$ and row~0 of $B$.

\definecolor{vframe}{gray}{0.62}\definecolor{vhead}{gray}{0.93}\definecolor{vink}{gray}{0.15}\definecolor{vlabel}{gray}{0.38}
\providecommand{\vfield}[1]{\textcolor{vlabel}{\textsc{#1}}}

\par\medskip\noindent\setlength{\fboxsep}{0pt}\setlength{\fboxrule}{0.8pt}\fcolorbox{vframe}{vhead}{\parbox{\dimexpr\linewidth-2\fboxrule\relax}{\small\vspace{3pt}\hspace{6pt}\textcolor{vink}{FOM objective, $C_{255}$, $(1,1)\times(1,1)$, \code{510}{40}{$\le$42}}\hspace{6pt}\vspace{3pt}}}\par\nointerlineskip
\vspace{-0.8pt}\noindent\fcolorbox{vframe}{white}{\begin{minipage}{\dimexpr\linewidth-2\fboxrule\relax}\small\vspace{3pt}\hspace{6pt}\begin{tabular}{@{}p{1.55cm}p{\dimexpr\linewidth-1.55cm-2\tabcolsep-12pt\relax}@{}}\vfield{Concept} & C255, cyclic order 255, shapes (1,1)x(1,1): two degree-eight character zeros plus a degree-two/four zero, with independently optimized intrinsic support-complexity pairings, should supply five fresh codes whose mean measured fitness exceeds cell 0.100's 74.6298. This is a falsifiable measured-score hypothesis, not a distance certificate. \\ \end{tabular}\vspace{3pt}\end{minipage}}\par\nointerlineskip
\vspace{-0.8pt}\noindent\fcolorbox{vframe}{white}{\begin{minipage}{\dimexpr\linewidth-2\fboxrule\relax}\small\vspace{3pt}\hspace{6pt}\begin{tabular}{@{}p{1.55cm}p{\dimexpr\linewidth-1.55cm-2\tabcolsep-12pt\relax}@{}}\vfield{Group} & L3 jump from C254 to genuinely different odd-order C255, using a one-coordinate cyclic presentation. This square bed is absent from BEDS. The square C254 parent scored 63.5622 with only six distinct codes; the different C17 x C17 square leader scored 74.6298. Since 255 divides 256-1, GF256 supplies characters with binary orbit degrees eight, four and two. Odd order makes F2[C255] semisimple. Every screened matrix action is compiled from verified unit-cell lifts and every emitted assembly is checked against a complete verified lift. \\ \end{tabular}\vspace{3pt}\end{minipage}}\par\nointerlineskip
\vspace{-0.8pt}\noindent\fcolorbox{vframe}{white}{\begin{minipage}{\dimexpr\linewidth-2\fboxrule\relax}\small\vspace{3pt}\hspace{6pt}\begin{tabular}{@{}p{1.55cm}p{\dimexpr\linewidth-1.55cm-2\tabcolsep-12pt\relax}@{}}\vfield{Shapes} & Keep equal (1,1) sides for open-ended cell 0.100: n=510, kb=0, guaranteed rate zero, equal 255-variable/255-check constituent maps, and shape lift ceiling 299. This group fixes order 255. The parent's minimum families touched both sector-one Acol0/Bcol0 and sector-two Arow0/Brow0; slot 8 fell from d2s=36 to d30s=24. Exact assembled ranks require 32$<$=k and 8*k$<$n, and logical screening includes both sectors and types. \\ \end{tabular}\vspace{3pt}\end{minipage}}\par\nointerlineskip
\vspace{-0.8pt}\noindent\fcolorbox{vframe}{white}{\begin{minipage}{\dimexpr\linewidth-2\fboxrule\relax}\small\vspace{3pt}\hspace{6pt}\begin{tabular}{@{}p{1.55cm}p{\dimexpr\linewidth-1.55cm-2\tabcolsep-12pt\relax}@{}}\vfield{Filling} & Twenty cards cross five common-character prescriptions with four intrinsic support-complexity pairings. Pair/triple character-syndrome matching enumerates fresh five-term cyclic supports, normalizes independent translation gauges, and excludes supports contained in proper cyclic cosets. Two-character relaxation is explicit only when a three-character library is too small; exact assembled ranks remain mandatory. Both CSS check weights are ten and every line is nonzero. Search ordinary and order-three-invariant physical supports with exact stabilizer classes, retain logical directions separately, use equal deeper finalist budgets and single/pair stabilizer dressings, and reject repeated uncoloured Tanner invariants. All found logical weights remain upper witnesses. \\ \end{tabular}\vspace{3pt}\end{minipage}}\par\nointerlineskip
\medskip

\par\medskip\noindent\setlength{\fboxsep}{0pt}\setlength{\fboxrule}{0.8pt}\fcolorbox{vframe}{vhead}{\parbox{\dimexpr\linewidth-2\fboxrule\relax}{\small\vspace{3pt}\hspace{6pt}\textcolor{vink}{CLER objective, $\mathrm{Dic}_{11}$, $(2,3)\times(2,3)$, \code{572}{67}{22}}\hspace{6pt}\vspace{3pt}}}\par\nointerlineskip
\vspace{-0.8pt}\noindent\fcolorbox{vframe}{white}{\begin{minipage}{\dimexpr\linewidth-2\fboxrule\relax}\small\vspace{3pt}\hspace{6pt}\begin{tabular}{@{}p{1.55cm}p{\dimexpr\linewidth-1.55cm-2\tabcolsep-12pt\relax}@{}}\vfield{Concept} & Dic11, order 44, shapes (2,3)x(2,3): asymmetric binomials with order-22 rotation differences in A and order-four mixed-coset differences in B should retain near-zero circuit failures while increasing measured k to 60-70 in the 0.100 cell. \\ \end{tabular}\vspace{3pt}\end{minipage}}\par\nointerlineskip
\vspace{-0.8pt}\noindent\fcolorbox{vframe}{white}{\begin{minipage}{\dimexpr\linewidth-2\fboxrule\relax}\small\vspace{3pt}\hspace{6pt}\begin{tabular}{@{}p{1.55cm}p{\dimexpr\linewidth-1.55cm-2\tabcolsep-12pt\relax}@{}}\vfield{Group} & Keep the frozen Dic11 bed: current fitness is 7.1162 and four measured codes had zero failures in both bases. Its C4 abelianization permits different binomial vanishing orders, motivating a filling-only rank-deficit experiment; a circuit advantage remains unverified. \\ \end{tabular}\vspace{3pt}\end{minipage}}\par\nointerlineskip
\vspace{-0.8pt}\noindent\fcolorbox{vframe}{white}{\begin{minipage}{\dimexpr\linewidth-2\fboxrule\relax}\small\vspace{3pt}\hspace{6pt}\begin{tabular}{@{}p{1.55cm}p{\dimexpr\linewidth-1.55cm-2\tabcolsep-12pt\relax}@{}}\vfield{Shapes} & Frozen equal (2,3) components have 132 variables and 88 checks each; n=572 and kb=44 give guaranteed rate 1/13. Exact ranks require 60$<$=k$<$=70, still below rate 0.125. Parent slot 34 had two Z failures, with Acol1/Bcol2 enriched 2.0x, but four measured siblings had none; this is insufficient evidence for a unique block repair. \\ \end{tabular}\vspace{3pt}\end{minipage}}\par\nointerlineskip
\vspace{-0.8pt}\noindent\fcolorbox{vframe}{white}{\begin{minipage}{\dimexpr\linewidth-2\fboxrule\relax}\small\vspace{3pt}\hspace{6pt}\begin{tabular}{@{}p{1.55cm}p{\dimexpr\linewidth-1.55cm-2\tabcolsep-12pt\relax}@{}}\vfield{Filling} & All cells are doubletons: component rows have weight six and columns weight four, so CSS check weight is ten and six-round CNOT count is 31680. Slots use canonical order-22 ratio architectures, not saved-code relabelings. A uses same-coset odd rotation displacements excluding the central involution; B uses mixed cosets. Verified lifts screen exact k, component words through weight four, cycles and small-boundary triples, then actual short quantum logicals and CSS. Sampled component-distance rewards saturate at 22. This targets the parent's ceiling limitation and rare collective Z failures, not distance as fitness. \\ \end{tabular}\vspace{3pt}\end{minipage}}\par\nointerlineskip
\medskip

\section{Protographs of the representative codes}\label{supp:protographs}
For each code of Table~I and Table~\ref{tab:w9cells}, we give its group by generators and relations, and the matrices $A$ and $B$.  Each entry of $A$ and $B$ is a sum of group elements, and $e$ is the identity.  Lifting $A$ and $B$ as in Supplemental Material~\ref{supp:lp} gives the code, up to the order of its qubits and checks.  The check matrices $H_X$ and $H_Z$ of every code are deposited in the repository~\cite{repo}.

\paragraph*{Weight-12 run, $k/n<0.125$, FOM objective (\code{510}{40}{$\le$42}, $G=C_{255}$).}
Generator $x$: \ $x^{255}=e$, $|G|=255$.
\begingroup\footnotesize
\[
A=\begin{pmatrix}
e+x^{2}+x^{38}+x^{103}+x^{193}
\end{pmatrix},\qquad B=\begin{pmatrix}
e+x^{22}+x^{46}+x^{156}+x^{227}
\end{pmatrix}.
\]
\endgroup

\paragraph*{Weight-12 run, $k/n<0.125$, FOM objective, second two-block code (\code{578}{32}{$\le$46}, $G=C_{17}\times C_{17}$).}
Generators $x,y$: \ $x^{17}=y^{17}=e,\ xy=yx$, $|G|=289$.
\begingroup\footnotesize
\[
A=\begin{pmatrix}
e+x^{11}y+x^{2}y^{7}+x^{14}y^{13}+x^{9}y^{14}
\end{pmatrix},\qquad B=\begin{pmatrix}
e+x^{5}y^{2}+x^{4}y^{9}+x^{11}y^{11}+xy^{14}
\end{pmatrix}.
\]
\endgroup

\paragraph*{Weight-12 run, $k/n<0.125$, CLER objective (\code{572}{67}{22}, $G=\mathrm{Dic}_{11}$, non-abelian).}
Generators $x,y$: \ $x^{22}=e,\ y^{2}=x^{11},\ yxy^{-1}=x^{-1}$, $|G|=44$.
\begingroup\footnotesize
\[
A=\begin{pmatrix}
x^{16}+x^{17} & x^{15}+x^{16} & x^{17}+x^{20} \\
x^{10}+x^{13} & x^{6}y+x^{15}y & x^{19}y+x^{20}y
\end{pmatrix},\qquad B=\begin{pmatrix}
x^{8}+x^{20}y & x^{21}+x^{8}y & x^{17}+x^{8}y \\
x^{6}+xy & x^{8}+x^{20}y & x^{21}+x^{5}y
\end{pmatrix}.
\]
\endgroup

\paragraph*{Weight-12 run, $k/n=0.125$--$0.175$, FOM objective (\code{592}{83}{18}, $G=C_{37}$).}
Generator $x$: \ $x^{37}=e$, $|G|=37$.
\begingroup\footnotesize
\[
A=\begin{pmatrix}
e+x^{8} & x+x^{36} & e+x^{25} \\
x^{14}+x^{32} & x^{10}+x^{34} & x^{3}+x^{13}
\end{pmatrix},\qquad B=\begin{pmatrix}
e+x^{25}+x^{34} & e+x^{10}+x^{21} & e & x^{36} \\
x^{20} & x^{31} & x^{26}+x^{29}+x^{30} & x^{2}+x^{10}+x^{34}
\end{pmatrix}.
\]
\endgroup

\paragraph*{Weight-12 run, $k/n=0.125$--$0.175$, CLER objective (\code{576}{82}{15}, $G=\mathrm{Dic}_{9}$, non-abelian).}
Generators $x,y$: \ $x^{18}=e,\ y^{2}=x^{9},\ yxy^{-1}=x^{-1}$, $|G|=36$.
\begingroup\footnotesize
\[
A=\begin{pmatrix}
x^{8}+x^{12} & x^{8}y+x^{10}y & x^{4}+x^{14} \\
x^{13}+x^{7}y & x^{15}+x^{4}y & x^{3}+x^{4}y
\end{pmatrix},\qquad B=\begin{pmatrix}
x^{8} & x^{12}+y & x^{17} & x^{17}+y+xy \\
y+x^{16}y & x^{4} & x^{17}+x^{2}y+x^{3}y & x^{6}
\end{pmatrix}.
\]
\endgroup

\paragraph*{Weight-12 run, $k/n=0.175$--$0.225$, FOM objective (\code{585}{121}{24}, $G=C_{13}\rtimes C_{9}$, non-abelian).}
Generators $x,y$: \ $x^{13}=y^{9}=e,\ yxy^{-1}=x^{3}$, $|G|=117$.
\begingroup\footnotesize
\[
A=\begin{pmatrix}
x^{3}+x^{5}y+x^{3}y^{5}+x^{8}y^{5} & x^{9}+x^{5}y^{2}+y^{3}+x^{3}y^{5}
\end{pmatrix},\qquad B=\begin{pmatrix}
x^{9}+x^{12}y^{2}+x^{7}y^{4}+x^{4}y^{5} & x^{11}y+xy^{2}+x^{7}y^{5}+x^{2}y^{7}
\end{pmatrix}.
\]
\endgroup

\paragraph*{Weight-12 run, $k/n=0.175$--$0.225$, CLER objective (\code{580}{116}{12}, $G=C_{29}$).}
Generator $x$: \ $x^{29}=e$, $|G|=29$.
\begingroup\footnotesize
\[
A=\begin{pmatrix}
x^{15} & x^{6} & x^{19}+x^{20} & x+x^{14} \\
e+x^{17} & x^{2}+x^{12} & x & x^{18}
\end{pmatrix},\qquad B=\begin{pmatrix}
x^{27} & x & x^{17}+x^{27} & x^{11}+x^{13} \\
x^{2}+x^{5} & x^{3}+x^{4} & x^{13} & x^{3}+x^{22}
\end{pmatrix}.
\]
\endgroup

\paragraph*{Weight-12 run, $k/n=0.225$--$0.275$, FOM objective (\code{560}{130}{14}, $G=\mathrm{Dic}_{7}$, non-abelian).}
Generators $x,y$: \ $x^{14}=e,\ y^{2}=x^{7},\ yxy^{-1}=x^{-1}$, $|G|=28$.
\begingroup\footnotesize
\[
A=\begin{pmatrix}
e+x^{6}y & e+x^{2}y & e+x^{5}y & e+y \\
x^{10}+x^{13}y & x+y & x^{8}+x^{2}y & x^{12}+x^{6}y
\end{pmatrix},\qquad B=\begin{pmatrix}
e+x^{11}y & e+y & e+x^{9} & e+x^{12}y \\
x^{4}+x^{4}y & e+x^{12}y & x^{10}+x^{10}y & x^{12}+x^{11}y
\end{pmatrix}.
\]
\endgroup

\paragraph*{Weight-12 run, $k/n=0.225$--$0.275$, CLER objective (\code{580}{132}{14}, $G=C_{29}$).}
Generator $x$: \ $x^{29}=e$, $|G|=29$.
\begingroup\footnotesize
\[
A=\begin{pmatrix}
x+x^{4} & x^{10}+x^{14} & x^{10}+x^{19} & x^{13}+x^{27} \\
x^{5}+x^{22} & x^{19}+x^{26} & x^{9}+x^{13} & x^{8}+x^{9}
\end{pmatrix},\qquad B=\begin{pmatrix}
x^{19}+x^{24} & x^{5}+x^{12} & x^{11}+x^{19} & x^{11}+x^{20} \\
x^{2}+x^{9} & x^{14}+x^{15} & x^{10}+x^{27} & x^{9}+x^{12}
\end{pmatrix}.
\]
\endgroup

\paragraph*{Weight-12 run, $k/n=0.275$--$0.325$, FOM objective (\code{588}{176}{12}, $G=\mathrm{Dic}_{21}$, non-abelian).}
Generators $x,y$: \ $x^{42}=e,\ y^{2}=x^{21},\ yxy^{-1}=x^{-1}$, $|G|=84$.
\begingroup\footnotesize
\[
A=\begin{pmatrix}
x^{20}+x^{10}y+x^{35}y & x^{33}+x^{10}y+x^{21}y
\end{pmatrix},\qquad B=\begin{pmatrix}
x^{14}+x^{33}+x^{5}y & x^{9}+x^{12}y+x^{17}y & x^{8}+x^{37}+x^{4}y
\end{pmatrix}.
\]
\endgroup

\paragraph*{Weight-12 run, $k/n=0.275$--$0.325$, CLER objective (\code{588}{176}{10}, $G=C_2\times(C_7\rtimes C_3)$, non-abelian).}
Generators $x,y,z$: \ $x^{7}=y^{3}=z^{2}=e,\ yxy^{-1}=x^{2},\ xz=zx,\ yz=zy$, $|G|=42$.
\begingroup\footnotesize
\[
A=\begin{pmatrix}
x^{5}y+x^{3}y^{2}+x^{5}z & x^{6}y^{2}+x^{6}z+x^{3}yz
\end{pmatrix},\qquad B=\begin{pmatrix}
x^{2}y & x^{5}yz & x^{4}z & x+x^{4}y^{2}z & xy^{2}z+x^{2}y^{2}z & x^{2}y^{2}+xz \\
x^{2}z+x^{5}z & x^{5}y^{2}+x^{6}z & x^{4}z+x^{5}z & x^{2}y^{2} & x^{4}z & x^{5}
\end{pmatrix}.
\]
\endgroup

\paragraph*{Weight-12 run, $k/n=0.325$--$0.375$, FOM objective (\code{544}{200}{8}, $G=Q_{16}$, non-abelian).}
Generators $x,y$: \ $x^{8}=e,\ y^{2}=x^{4},\ yxy^{-1}=x^{-1}$, $|G|=16$.
\begingroup\footnotesize
\[
A=\begin{pmatrix}
e+x^{2} & e+y & e+x^{2}y & e+y & 0 \\
x^{2}y & x^{7} & x^{3} & x^{2} & x+x^{4}+xy
\end{pmatrix},\qquad B=\begin{pmatrix}
x^{5}+x^{6}y & x^{2}+y & y+x^{2}y & x^{6}y & x^{7} & e \\
x^{7}y & x^{5} & x^{6} & x^{2}y+x^{3}y & x^{5}+x^{4}y & x^{3}+x^{5}y
\end{pmatrix}.
\]
\endgroup

\paragraph*{Weight-12 run, $k/n=0.325$--$0.375$, CLER objective (\code{598}{195}{7}, $G=C_{13}$).}
Generator $x$: \ $x^{13}=e$, $|G|=13$.
\begingroup\footnotesize
\[
A=\begin{pmatrix}
e & x^{9}+x^{10} & x+x^{3} & e & e+x^{10} \\
e & e & e & x^{6}+x^{7} & x^{5}+x^{11}
\end{pmatrix},\qquad B=\begin{pmatrix}
0 & 0 & e & e & e & e & e+x^{6} & e+x^{10} \\
e & e & 0 & x^{9} & x^{6} & x^{3}+x^{8} & x^{11} & x^{4} \\
x^{11} & x^{8}+x^{10} & x^{4}+x^{9} & x^{3} & x^{7} & 0 & 0 & 0
\end{pmatrix}.
\]
\endgroup

\paragraph*{Weight-12 run, $k/n=0.375$--$0.425$, FOM objective (\code{560}{224}{10}, $G=\mathrm{AGL}(1,8)$, non-abelian).}
Generators $t_0,t_1,t_2,y$: \ $t_i^{2}=e,\ t_it_j=t_jt_i,\ y^{7}=e,\ yt_0y^{-1}=t_1,\ yt_1y^{-1}=t_2,\ yt_2y^{-1}=t_0t_1$, $|G|=56$.
\begingroup\footnotesize
\[
A=\begin{pmatrix}
e+t_2y+t_1t_2y^{2} & e+t_0y+t_0t_1y^{4} & e+t_0t_1y^{2}+t_1t_2y^{4}
\end{pmatrix},\qquad B=\begin{pmatrix}
e+t_0t_2y+t_0t_1y^{3} & e+t_0t_2y^{2}+y^{4} & e+t_1y+t_0t_1t_2y^{2}
\end{pmatrix}.
\]
\endgroup

\paragraph*{Weight-12 run, $k/n=0.375$--$0.425$, CLER objective (\code{550}{220}{7}, $G=C_{11}\rtimes C_{5}$, non-abelian).}
Generators $x,y$: \ $x^{11}=y^{5}=e,\ yxy^{-1}=x^{3}$, $|G|=55$.
\begingroup\footnotesize
\[
A=\begin{pmatrix}
e+x & e+y+x^{4}y^{2} & e+x^{2}y+x^{9}y^{3}
\end{pmatrix},\qquad B=\begin{pmatrix}
e+x & e+y+x^{4}y^{2} & e+x^{9}y+x^{3}y^{3}
\end{pmatrix}.
\]
\endgroup

\paragraph*{Weight-9 run, $k/n<0.125$, FOM objective (\code{572}{48}{24}, $G=\mathrm{Dic}_{11}$, non-abelian).}
Generators $x,y$: \ $x^{22}=e,\ y^{2}=x^{11},\ yxy^{-1}=x^{-1}$, $|G|=44$.
\begingroup\footnotesize
\[
A=\begin{pmatrix}
x^{21}y & x & x^{16}+x^{18}+y \\
x^{11}y & x^{4}+x^{9}+x^{16}y & x^{9}
\end{pmatrix},\qquad B=\begin{pmatrix}
x^{18}y & x^{18} & x^{20}+x^{2}y+x^{15}y \\
y & x^{15}+x^{18}+x^{15}y & x^{14}y
\end{pmatrix}.
\]
\endgroup

\paragraph*{Weight-9 run, $k/n=0.125$--$0.175$, FOM objective (\code{496}{72}{14}, $G=C_{31}$).}
Generator $x$: \ $x^{31}=e$, $|G|=31$.
\begingroup\footnotesize
\[
A=\begin{pmatrix}
x^{5} & x^{6} & x^{4}+x^{8} \\
x^{18}+x^{19} & x^{4}+x^{9} & x^{14}
\end{pmatrix},\qquad B=\begin{pmatrix}
x^{3} & x^{20} & x+x^{2} & x^{2}+x^{5} \\
x^{19}+x^{28} & e+x^{26} & x^{10} & x
\end{pmatrix}.
\]
\endgroup

\paragraph*{Weight-9 run, $k/n=0.175$--$0.225$, FOM objective (\code{585}{119}{18}, $G=C_{13}\rtimes C_{9}$, non-abelian).}
Generators $x,y$: \ $x^{13}=y^{9}=e,\ yxy^{-1}=x^{3}$, $|G|=117$.
\begingroup\footnotesize
\[
A=\begin{pmatrix}
x^{9}y^{2}+x^{9}y^{3}+x^{10}y^{7} & x^{3}+x^{3}y^{2}+y^{7}
\end{pmatrix},\qquad B=\begin{pmatrix}
x^{4}y+x^{7}y^{2}+x^{6}y^{7} & x^{5}y^{2}+x^{9}y^{7}+x^{8}y^{8}
\end{pmatrix}.
\]
\endgroup

\paragraph*{Weight-9 run, $k/n=0.225$--$0.275$, FOM objective (\code{576}{150}{9}, $G=C_{3}\rtimes C_{16}$, non-abelian).}
Generators $x,y$: \ $x^{3}=y^{16}=e,\ yxy^{-1}=x^{2}$, $|G|=48$.
\begingroup\footnotesize
\[
A=\begin{pmatrix}
xy^{2}+xy^{9}+x^{2}y^{11} & x^{2}y^{7}+xy^{9}+y^{15}
\end{pmatrix},\qquad B=\begin{pmatrix}
e+x^{2}y & y^{5} & e & e & e \\
x^{2}y^{2} & xy^{7}+x^{2}y^{14} & e & y^{5} & xy^{12}
\end{pmatrix}.
\]
\endgroup

\paragraph*{Weight-9 run, $k/n=0.275$--$0.325$, FOM objective (\code{580}{180}{7}, $G=C_{5}\rtimes C_{4}$, non-abelian).}
Generators $x,y$: \ $x^{5}=y^{4}=e,\ yxy^{-1}=x^{2}$, $|G|=20$.
\begingroup\footnotesize
\[
A=\begin{pmatrix}
y+x^{2}y & x^{3}y^{2} & e & e & e \\
y^{2} & e+x^{3}y & e & x^{2} & xy^{3}
\end{pmatrix},\qquad B=\begin{pmatrix}
y+x^{2}y & x^{3}y^{2} & e & e & e \\
y^{2} & e+x^{3}y & e & x^{2} & xy^{3}
\end{pmatrix}.
\]
\endgroup

\paragraph*{Weight-9 run, $k/n=0.325$--$0.375$, FOM objective (\code{584}{200}{6}, $G=C_{8}$).}
Generator $x$: \ $x^{8}=e$, $|G|=8$.
\begingroup\footnotesize
\[
A=\begin{pmatrix}
e & e & 0 & 0 & 0 & e+x^{5} & e+x^{6} & 0 \\
0 & e & e & e & e & 0 & 0 & e+x^{5} \\
0 & 0 & e & x & x^{3} & x^{2} & x^{4} & x^{7}
\end{pmatrix},\qquad B=\begin{pmatrix}
e & e & 0 & 0 & 0 & e+x^{6} & e+x & 0 \\
0 & e & e & e & e & 0 & 0 & e+x^{5} \\
0 & 0 & e & x & x^{3} & x^{2} & x^{7} & x^{7}
\end{pmatrix}.
\]
\endgroup

\paragraph*{Weight-9 run, $k/n=0.375$--$0.425$, FOM objective (\code{550}{226}{5}, $G=C_{11}\rtimes C_{5}$, non-abelian).}
Generators $x,y$: \ $x^{11}=y^{5}=e,\ yxy^{-1}=x^{3}$, $|G|=55$.
\begingroup\footnotesize
\[
A=\begin{pmatrix}
e+x & e+y & e+x^{5}y^{2}
\end{pmatrix},\qquad B=\begin{pmatrix}
e+x & e+y^{2} & e+x^{5}y^{2}
\end{pmatrix}.
\]
\endgroup

\paragraph*{Weight-9 run, $k/n=0.425$--$0.475$, FOM objective (\code{552}{258}{4}, $G=C_{12}\rtimes C_{2}$, non-abelian).}
Generators $x,y$: \ $x^{12}=y^{2}=e,\ yxy^{-1}=x^{5}$, $|G|=24$.
\begingroup\footnotesize
\[
A=\begin{pmatrix}
e+x & e+x^{3} & e+x^{5}
\end{pmatrix},\qquad B=\begin{pmatrix}
e & e & e & e & e & e & e \\
e & x & x^{2} & x^{3} & x^{6} & x^{7} & x^{9}
\end{pmatrix}.
\]
\endgroup

\paragraph*{Weight-9 run, $k/n=0.475$--$0.525$, FOM objective (\code{583}{287}{4}, $G=C_{11}$).}
Generator $x$: \ $x^{11}=e$, $|G|=11$.
\begingroup\footnotesize
\[
A=\begin{pmatrix}
e+x & e+x^{3} & e & e & e & 0 & 0 \\
0 & 0 & e & x^{2} & x^{4} & e+x & e+x^{3}
\end{pmatrix},\qquad B=\begin{pmatrix}
e+x & e+x^{3} & e & e & e & 0 & 0 \\
0 & 0 & e & x^{2} & x^{4} & e+x & e+x^{3}
\end{pmatrix}.
\]
\endgroup

\paragraph*{Weight-9 run, $k/n>0.525$, FOM objective (\code{476}{252}{3}, $G=\mathrm{Dic}_{7}$, non-abelian).}
Generators $x,y$: \ $x^{14}=e,\ y^{2}=x^{7},\ yxy^{-1}=x^{-1}$, $|G|=28$.
\begingroup\footnotesize
\[
A=\begin{pmatrix}
e & x^{5}+x^{11}y & x+x^{3}y & x^{9}+x^{12}y
\end{pmatrix},\qquad B=\begin{pmatrix}
e & x^{3}+x^{8}y & x^{7}+x^{6}y & x^{5}+y
\end{pmatrix}.
\]
\endgroup

\paragraph*{Weight-9 run, $k/n=0.175$--$0.225$, FOM objective, code marked R (\code{585}{117}{20}, $G=C_{13}\rtimes C_{9}$, non-abelian).}
Generators $x,y$: \ $x^{13}=y^{9}=e,\ yxy^{-1}=x^{3}$, $|G|=117$.
\begingroup\footnotesize
\[
A=\begin{pmatrix}
x^{5}y+x^{4}y^{4}+x^{12}y^{6} & x^{3}+x^{10}y^{2}+y^{5}
\end{pmatrix},\qquad B=\begin{pmatrix}
x^{10}y+x^{9}y^{5}+x^{3}y^{7} & x^{7}+x^{9}y^{3}+x^{4}y^{5}
\end{pmatrix}.
\]
\endgroup

\section{Elite codes of every run}\label{supp:elites}
Tables~\ref{tab:elites-fom} and~\ref{tab:elites-cler} list the elite code of every rate cell for every
language-model run of Fig.~3, in the layout of Table~I.
Tables~\ref{tab:runcells-dist} and~\ref{tab:runcells-circ} give the best program score in every rate
cell for every run of Fig.~3, including the random-mutation and hill-climbing runs. The group names in these tables are those written by the programs. $C_n$ is the cyclic group of order $n$, and $\times$ denotes a direct product. $D_m$ is the dihedral group of order $2m$, and $\mathrm{Dic}_m=\langle x,y\mid x^{2m}=e,\ y^2=x^m,\ yxy^{-1}=x^{-1}\rangle$ is the dicyclic group of order $4m$, which includes the quaternion groups $Q_8=\mathrm{Dic}_2$ and $Q_{16}=\mathrm{Dic}_4$. $C_m\rtimes C_n$ is a semidirect product in which $C_m=\langle x\rangle$ is normal and a generator $y$ of $C_n$ acts by $yxy^{-1}=x^{r}$. Where different $r$ give different groups, the subscript gives $r$, as in $C_8\rtimes_3 C_4$. The remaining names are standard: $S_3$ and $A_4$ are the symmetric and alternating groups of orders $6$ and $12$, $\mathrm{SD}_{32}$ is the semidihedral group of order $32$, $\mathrm{SL}(2,3)$ is the special linear group over $\mathbb F_3$, $\mathrm{He}_3$ is the Heisenberg group over $\mathbb F_3$, and $\mathrm{AGL}(1,8)$ is the affine group of $\mathbb F_8$.

\begin{table*}[!htbp]
\centering
\scriptsize
\setlength{\tabcolsep}{4pt}
\caption{Elite codes of the FOM runs of Fig.~3, in the layout of Table~I.  For every rate cell and language-model run, the listed code is the best, by $kd^2/n$, of the five measured codes of the elite program of the cell.  The distances of the gpt-6.1-sol codes are certified (Supplemental Material~\ref{supp:exactd}), except for the two two-block codes of cell~0, whose distances are upper bounds from two-hour searches ($\dagger$).  For gpt-6.1-sol, cell~0 also lists the second two-block program of Table~I.  All other distances are thirty-second in-loop upper bounds ($\le$).  The column $w$ gives the check weight.  }
\label{tab:elites-fom}
\begin{tabular}{lllllc}
\toprule
Cell ($k/n$) & Model (effort) & Group & $A\times B$ & $[[n,k,d]]$ & $w$ \\
\midrule
0 ($<0.125$) & gpt-6-luna, low & $C_{46}$ & $(2,3)\times(2,3)$ & \code{598}{62}{$\le$23} & 10 \\
 & gpt-6-luna, high & $C_{46}$ & $(2,3)\times(2,3)$ & \code{598}{46}{$\le$28} & 12 \\
 & gpt-6-luna, extra-high & $C_{46}$ & $(2,3)\times(2,3)$ & \code{598}{62}{$\le$24} & 10 \\
 & gpt-6-sol, high & $C_{46}$ & $(3,2)\times(3,2)$ & \code{598}{46}{$\le$28} & 12 \\
 & gpt-6.1-sol, high & $C_{255}$ & $(1,1)\times(1,1)$ & \code{510}{40}{$\le$42$^{\dagger}$} & 10 \\
 & gpt-6.1-sol, high & $C_{17}\times C_{17}$ & $(1,1)\times(1,1)$ & \code{578}{32}{$\le$46$^{\dagger}$} & 10 \\
\addlinespace[2pt]
1 (0.15) & gpt-6-luna, low & $S_3\times C_9$ & $(1,2)\times(2,3)$ & \code{432}{57}{$\le$12} & 10 \\
 & gpt-6-luna, high & $C_8\rtimes_3 C_4$ & $(2,3)\times(2,4)$ & \code{512}{76}{$\le$12} & 10 \\
 & gpt-6-luna, extra-high & $C_{23}$ & $(2,3)\times(3,6)$ & \code{552}{90}{$\le$15} & 12 \\
 & gpt-6-sol, high & $C_{37}$ & $(2,3)\times(2,4)$ & \code{592}{83}{$\le$18} & 12 \\
 & gpt-6.1-sol, high & $C_{37}$ & $(2,3)\times(2,4)$ & \code{592}{83}{18} & 12 \\
\addlinespace[2pt]
2 (0.20) & gpt-6-luna, low & $S_3\times C_{19}$ & $(1,2)\times(1,2)$ & \code{570}{122}{$\le$21} & 12 \\
 & gpt-6-luna, high & $C_{11}\times D_5$ & $(1,2)\times(1,2)$ & \code{550}{118}{$\le$22} & 12 \\
 & gpt-6-luna, extra-high & $C_{13}\rtimes C_9$ & $(1,2)\times(1,2)$ & \code{585}{121}{$\le$24} & 12 \\
 & gpt-6-sol, high & $C_{13}\rtimes C_9$ & $(1,2)\times(1,2)$ & \code{585}{121}{$\le$24} & 12 \\
 & gpt-6.1-sol, high & $C_{13}\rtimes C_9$ & $(1,2)\times(1,2)$ & \code{585}{121}{24} & 12 \\
\addlinespace[2pt]
3 (0.25) & gpt-6-luna, low & $Q_8\times C_3$ & $(5,2)\times(4,2)$ & \code{576}{144}{$\le$10} & 11 \\
 & gpt-6-luna, high & $C_{14}\times C_2$ & $(2,4)\times(2,4)$ & \code{560}{134}{$\le$12} & 12 \\
 & gpt-6-luna, extra-high & $C_{13}\rtimes C_9$ & $(2,1)\times(2,1)$ & \code{585}{153}{$\le$13} & 12 \\
 & gpt-6-sol, high & $\mathrm{Dic}_{12}$ & $(1,2)\times(2,5)$ & \code{576}{156}{$\le$12} & 12 \\
 & gpt-6.1-sol, high & $\mathrm{Dic}_{7}$ & $(2,4)\times(2,4)$ & \code{560}{130}{14} & 12 \\
\addlinespace[2pt]
4 (0.30) & gpt-6-luna, low & $C_{19}$ & $(2,5)\times(2,5)$ & \code{551}{171}{$\le$7} & 10 \\
 & gpt-6-luna, high & $S_3\times C_7$ & $(1,3)\times(2,4)$ & \code{588}{178}{$\le$7} & 12 \\
 & gpt-6-luna, extra-high & $C_7\rtimes C_{16}$ & $(2,1)\times(2,1)$ & \code{560}{176}{$\le$7} & 12 \\
 & gpt-6-sol, high & $C_{19}$ & $(2,5)\times(2,5)$ & \code{551}{171}{$\le$9} & 12 \\
 & gpt-6.1-sol, high & $\mathrm{Dic}_{21}$ & $(1,2)\times(1,3)$ & \code{588}{176}{12} & 12 \\
\addlinespace[2pt]
5 (0.35) & gpt-6-luna, low & $C_{17}$ & $(2,5)\times(2,6)$ & \code{578}{204}{$\le$6} & 12 \\
 & gpt-6-luna, high & $C_{17}$ & $(2,5)\times(2,6)$ & \code{578}{204}{$\le$7} & 12 \\
 & gpt-6-luna, extra-high & $C_{17}$ & $(2,5)\times(2,6)$ & \code{578}{204}{$\le$7} & 12 \\
 & gpt-6-sol, high & $C_8\rtimes_5 C_4$ & $(1,3)\times(2,5)$ & \code{544}{200}{$\le$8} & 12 \\
 & gpt-6.1-sol, high & $Q_{16}$ & $(2,5)\times(2,6)$ & \code{544}{200}{8} & 12 \\
\addlinespace[2pt]
6 (0.40) & gpt-6-luna, low & $C_{14}$ & $(6,2)\times(6,2)$ & \code{560}{224}{$\le$6} & 12 \\
 & gpt-6-luna, high & $S_3\times C_9$ & $(1,3)\times(1,3)$ & \code{540}{220}{$\le$8} & 12 \\
 & gpt-6-luna, extra-high & $D_{7}$ & $(6,2)\times(6,2)$ & \code{560}{224}{$\le$6} & 12 \\
 & gpt-6-sol, high & $C_{11}\rtimes C_5$ & $(1,3)\times(1,3)$ & \code{550}{220}{$\le$10} & 12 \\
 & gpt-6.1-sol, high & $\mathrm{AGL}(1,8)$ & $(1,3)\times(1,3)$ & \code{560}{224}{10} & 12 \\
\addlinespace[2pt]
7 (0.45) & gpt-6-luna, low & $S_3\times C_7$ & $(1,3)\times(1,4)$ & \code{546}{260}{$\le$4} & 11 \\
 & gpt-6-luna, high & $S_3\times C_6$ & $(1,4)\times(1,3)$ & \code{468}{222}{$\le$4} & 12 \\
 & gpt-6-luna, extra-high & $C_{46}$ & $(1,3)\times(1,4)$ & \code{598}{283}{$\le$4} & 10 \\
 & gpt-6-sol, high & $Q_8\times C_3$ & $(1,3)\times(2,7)$ & \code{552}{250}{$\le$6} & 12 \\
 & gpt-6.1-sol, high & $C_7\rtimes C_3$ & $(1,3)\times(2,7)$ & \code{483}{220}{6} & 12 \\
\addlinespace[2pt]
8 (0.50) & gpt-6-luna, low & $C_{42}$ & $(1,3)\times(1,4)$ & \code{546}{263}{$\le$4} & 11 \\
 & gpt-6-luna, high & $C_6\times C_3$ & $(1,4)\times(2,7)$ & \code{540}{275}{$\le$4} & 12 \\
 & gpt-6-luna, extra-high & $C_{17}$ & $(1,4)\times(2,7)$ & \code{510}{258}{$\le$4} & 12 \\
 & gpt-6-sol, high & $\mathrm{Dic}_{9}$ & $(1,3)\times(1,5)$ & \code{576}{300}{$\le$4} & 12 \\
 & gpt-6.1-sol, high & $C_5\rtimes C_8$ & $(1,3)\times(1,4)$ & \code{520}{252}{5} & 12 \\
\addlinespace[2pt]
9 ($>0.525$) & gpt-6-luna, low & $C_{28}$ & $(1,4)\times(1,5)$ & \code{588}{345}{$\le$3} & 12 \\
 & gpt-6-luna, high & $Q_{16}\times C_2$ & $(1,4)\times(1,4)$ & \code{544}{294}{$\le$4} & 12 \\
 & gpt-6-luna, extra-high & $\mathrm{Dic}_{8}$ & $(1,4)\times(1,4)$ & \code{544}{296}{$\le$4} & 12 \\
 & gpt-6-sol, high & $\mathrm{Dic}_{7}$ & $(1,4)\times(1,5)$ & \code{588}{345}{$\le$4} & 12 \\
 & gpt-6.1-sol, high & $S_3\times C_4$ & $(1,4)\times(1,5)$ & \code{504}{297}{4} & 12 \\
\bottomrule
\end{tabular}
\end{table*}

\begin{table*}[!htbp]
\centering
\scriptsize
\setlength{\tabcolsep}{4pt}
\caption{As Table~\ref{tab:elites-fom}, for the CLER runs.  The listed code is the best of the five measured codes by its term of Eq.~(3). The last column gives its logical error rate per round and per logical qubit at $p=0.002$, as measured during the run in the basis with the larger rate (up to $2\times10^5$ shots or $100$ failures). For codes without failures, it gives the upper end of the two-sided $95\%$ Clopper--Pearson interval ($<$).  The distances of the gpt-6.1-sol codes are certified (Section~\ref{supp:exactd}).  All other distances are two-second in-loop upper bounds ($\le$).}
\label{tab:elites-cler}
\begin{tabular}{lllllcr}
\toprule
Cell ($k/n$) & Model (effort) & Group & $A\times B$ & $[[n,k,d]]$ & $w$ & $p_{\mathrm L}$ at $p=0.002$ \\
\midrule
0 ($<0.125$) & gpt-6-luna, low & $Q_8\times C_5$ & $(2,3)\times(2,3)$ & \code{520}{55}{$\le$20} & 10 & $<5.6\times10^{-8}$ \\
 & gpt-6-luna, high & $C_{15}\times C_3$ & $(2,3)\times(2,3)$ & \code{585}{61}{$\le$15} & 10 & $<5.0\times10^{-8}$ \\
 & gpt-6-luna, extra-high & $C_{46}$ & $(2,3)\times(2,3)$ & \code{598}{46}{$\le$14} & 8 & $<6.7\times10^{-8}$ \\
 & gpt-6-sol, high & $C_{46}$ & $(2,3)\times(2,3)$ & \code{598}{70}{$\le$23} & 10 & $<4.4\times10^{-8}$ \\
 & gpt-6.1-sol, high & $\mathrm{Dic}_{11}$ & $(2,3)\times(2,3)$ & \code{572}{67}{22} & 10 & $<4.6\times10^{-8}$ \\
\addlinespace[2pt]
1 (0.15) & gpt-6-luna, low & $C_{28}$ & $(2,3)\times(2,4)$ & \code{448}{78}{$\le$12} & 12 & $6.4\times10^{-8}$ \\
 & gpt-6-luna, high & $C_{23}$ & $(2,4)\times(3,5)$ & \code{598}{92}{$\le$10} & 10 & $7.2\times10^{-8}$ \\
 & gpt-6-luna, extra-high & $C_{23}$ & $(2,4)\times(3,5)$ & \code{598}{94}{$\le$10} & 11 & $1.8\times10^{-8}$ \\
 & gpt-6-sol, high & $C_{37}$ & $(2,3)\times(2,4)$ & \code{592}{88}{$\le$16} & 12 & $<3.5\times10^{-8}$ \\
 & gpt-6.1-sol, high & $\mathrm{Dic}_{9}$ & $(2,3)\times(2,4)$ & \code{576}{82}{15} & 11 & $<3.7\times10^{-8}$ \\
\addlinespace[2pt]
2 (0.20) & gpt-6-luna, low & $\mathrm{Dic}_{7}$ & $(2,4)\times(2,4)$ & \code{560}{112}{$\le$9} & 10 & $<2.7\times10^{-8}$ \\
 & gpt-6-luna, high & $C_9\times C_3$ & $(2,4)\times(2,4)$ & \code{540}{108}{$\le$10} & 9 & $7.7\times10^{-9}$ \\
 & gpt-6-luna, extra-high & $\mathrm{Dic}_{7}$ & $(2,4)\times(2,4)$ & \code{560}{112}{$\le$9} & 9 & $<2.7\times10^{-8}$ \\
 & gpt-6-sol, high & $C_{29}$ & $(2,4)\times(2,4)$ & \code{580}{122}{$\le$12} & 11 & $<2.5\times10^{-8}$ \\
 & gpt-6.1-sol, high & $C_{29}$ & $(2,4)\times(2,4)$ & \code{580}{116}{12} & 10 & $<2.7\times10^{-8}$ \\
\addlinespace[2pt]
3 (0.25) & gpt-6-luna, low & $C_{28}$ & $(2,4)\times(2,4)$ & \code{560}{128}{$\le$12} & 12 & $5.9\times10^{-8}$ \\
 & gpt-6-luna, high & $C_{24}$ & $(5,2)\times(4,2)$ & \code{576}{144}{$\le$9} & 12 & $1.8\times10^{-7}$ \\
 & gpt-6-luna, extra-high & $\mathrm{Dic}_{6}$ & $(5,2)\times(4,2)$ & \code{576}{144}{$\le$9} & 11 & $2.9\times10^{-7}$ \\
 & gpt-6-sol, high & $C_{29}$ & $(2,4)\times(2,4)$ & \code{580}{132}{$\le$13} & 12 & $1.3\times10^{-8}$ \\
 & gpt-6.1-sol, high & $C_{29}$ & $(2,4)\times(2,4)$ & \code{580}{132}{14} & 12 & $1.3\times10^{-8}$ \\
\addlinespace[2pt]
4 (0.30) & gpt-6-luna, low & $C_{17}$ & $(2,5)\times(2,5)$ & \code{493}{153}{$\le$7} & 12 & $5.7\times10^{-6}$ \\
 & gpt-6-luna, high & $C_{10}\times C_2$ & $(2,5)\times(2,5)$ & \code{580}{180}{$\le$7} & 10 & $1.8\times10^{-6}$ \\
 & gpt-6-luna, extra-high & $D_{9}$ & $(2,5)\times(2,5)$ & \code{522}{162}{$\le$6} & 10 & $9.0\times10^{-6}$ \\
 & gpt-6-sol, high & $C_7\times A_4$ & $(1,2)\times(1,3)$ & \code{588}{168}{$\le$10} & 12 & $1.6\times10^{-7}$ \\
 & gpt-6.1-sol, high & $C_2\times(C_7\rtimes C_3)$ & $(1,2)\times(2,6)$ & \code{588}{176}{10} & 12 & $2.8\times10^{-7}$ \\
\addlinespace[2pt]
5 (0.35) & gpt-6-luna, low & $C_{17}$ & $(2,5)\times(2,6)$ & \code{578}{204}{$\le$6} & 12 & $1.9\times10^{-5}$ \\
 & gpt-6-luna, high & $C_{17}$ & $(2,5)\times(2,6)$ & \code{578}{204}{$\le$6} & 12 & $1.3\times10^{-5}$ \\
 & gpt-6-luna, extra-high & $C_{17}$ & $(2,5)\times(2,6)$ & \code{578}{204}{$\le$6} & 11 & $1.6\times10^{-5}$ \\
 & gpt-6-sol, high & $C_{20}$ & $(2,5)\times(2,5)$ & \code{580}{200}{$\le$7} & 12 & $7.1\times10^{-6}$ \\
 & gpt-6.1-sol, high & $C_{13}$ & $(2,5)\times(3,8)$ & \code{598}{195}{7} & 12 & $4.1\times10^{-6}$ \\
\addlinespace[2pt]
6 (0.40) & gpt-6-luna, low & $C_{14}$ & $(6,2)\times(6,2)$ & \code{560}{224}{$\le$5} & 12 & $4.8\times10^{-5}$ \\
 & gpt-6-luna, high & $D_{7}$ & $(6,2)\times(6,2)$ & \code{560}{224}{$\le$5} & 12 & $3.1\times10^{-5}$ \\
 & gpt-6-luna, extra-high & $C_{19}\rtimes C_3$ & $(1,3)\times(1,3)$ & \code{570}{228}{$\le$8} & 12 & $9.1\times10^{-6}$ \\
 & gpt-6-sol, high & $C_7\rtimes C_8$ & $(1,3)\times(1,3)$ & \code{560}{224}{$\le$8} & 12 & $7.7\times10^{-6}$ \\
 & gpt-6.1-sol, high & $C_{11}\rtimes C_5$ & $(1,3)\times(1,3)$ & \code{550}{220}{7} & 11 & $4.4\times10^{-6}$ \\
\addlinespace[2pt]
7 (0.45) & gpt-6-luna, low & $S_3\times C_5$ & $(1,3)\times(1,4)$ & \code{390}{182}{$\le$3} & 11 & $4.6\times10^{-4}$ \\
 & gpt-6-luna, high & $\mathrm{SL}(2,3)$ & $(1,3)\times(2,7)$ & \code{552}{240}{$\le$4} & 12 & $9.4\times10^{-5}$ \\
 & gpt-6-luna, extra-high & $Q_8\times C_3$ & $(1,3)\times(2,7)$ & \code{552}{240}{$\le$5} & 12 & $6.5\times10^{-5}$ \\
 & gpt-6-sol, high & $C_{26}$ & $(1,3)\times(2,7)$ & \code{598}{260}{$\le$6} & 12 & $4.7\times10^{-5}$ \\
 & gpt-6.1-sol, high & $C_3\rtimes C_8$ & $(1,3)\times(2,7)$ & \code{552}{240}{5} & 12 & $3.9\times10^{-5}$ \\
\addlinespace[2pt]
8 (0.50) & gpt-6-luna, low & $S_3\times C_5$ & $(1,3)\times(1,4)$ & \code{390}{187}{$\le$3} & 10 & $3.8\times10^{-4}$ \\
 & gpt-6-luna, high & $C_5\rtimes C_8$ & $(1,3)\times(1,4)$ & \code{520}{247}{$\le$4} & 10 & $2.5\times10^{-4}$ \\
 & gpt-6-luna, extra-high & $C_{23}$ & $(1,3)\times(1,5)$ & \code{368}{184}{$\le$3} & 12 & $6.1\times10^{-4}$ \\
 & gpt-6-sol, high & $C_{19}$ & $(1,4)\times(2,7)$ & \code{570}{288}{$\le$4} & 12 & $3.2\times10^{-4}$ \\
 & gpt-6.1-sol, high & $C_{19}$ & $(1,4)\times(2,7)$ & \code{570}{288}{4} & 12 & $3.3\times10^{-4}$ \\
\addlinespace[2pt]
9 ($>0.525$) & gpt-6-luna, low & $\mathrm{He}_3$ & $(1,4)\times(1,4)$ & \code{459}{251}{$\le$3} & 10 & $4.3\times10^{-4}$ \\
 & gpt-6-luna, high & $C_{31}$ & $(4,1)\times(4,1)$ & \code{527}{287}{$\le$3} & 10 & $3.6\times10^{-4}$ \\
 & gpt-6-luna, extra-high & $C_{33}$ & $(1,4)\times(1,4)$ & \code{561}{297}{$\le$4} & 12 & $2.7\times10^{-4}$ \\
 & gpt-6-sol, high & $C_{34}$ & $(1,4)\times(1,4)$ & \code{578}{306}{$\le$4} & 12 & $2.5\times10^{-4}$ \\
 & gpt-6.1-sol, high & $\mathrm{SD}_{32}$ & $(1,4)\times(1,4)$ & \code{544}{288}{4} & 12 & $2.2\times10^{-4}$ \\
\bottomrule
\end{tabular}
\end{table*}

\begin{table*}[!htbp]
\centering
\footnotesize
\setlength{\tabcolsep}{3.5pt}
\caption{Best in-loop program score $\mathcal{S}_{\mathrm{FOM}}$ in every rate cell for every FOM run of Fig.~3, after the eight seed programs and 500 steps.  Columns 0--9 are the rate cells, whose centres run from $k/n=0.10$ to $0.55$.  The last column is the run score.}
\label{tab:runcells-dist}
\begin{tabular}{lrrrrrrrrrrr}
\toprule
Run & 0 & 1 & 2 & 3 & 4 & 5 & 6 & 7 & 8 & 9 & Run score \\
\midrule
Random mutation, repeat 1 & 28.2 & 23.4 & 35.8 & 14.7 & 11.3 & 11.2 & 10.0 & 7.1 & 4.5 & 3.4 & 149.5 \\
Random mutation, repeat 2 & 28.0 & 25.4 & 35.8 & 14.7 & 11.3 & 12.1 & 10.2 & 5.3 & 7.7 & 4.8 & 155.1 \\
Random mutation, repeat 3 & 21.9 & 17.8 & 32.1 & 15.1 & 11.4 & 11.2 & 9.5 & 6.8 & 6.6 & 3.0 & 135.4 \\
Random mutation, repeat 4 & 38.9 & 23.9 & 33.1 & 14.7 & 12.0 & 11.2 & 8.6 & 6.4 & 7.6 & 2.9 & 159.3 \\
Random mutation, repeat 5 & 39.5 & 18.6 & 30.9 & 16.0 & 11.4 & 12.2 & 7.3 & 7.1 & 3.5 & 3.0 & 149.6 \\
Hill-climbing, repeat 1 & 27.0 & 22.2 & 54.4 & 20.7 & 11.4 & 12.7 & 25.6 & 7.6 & 7.7 & 4.8 & 194.0 \\
Hill-climbing, repeat 2 & 25.5 & 22.6 & 73.7 & 20.4 & 19.9 & 17.3 & 15.4 & 7.6 & 5.4 & 4.9 & 212.8 \\
Hill-climbing, repeat 3 & 54.3 & 22.6 & 64.8 & 20.4 & 11.6 & 15.5 & 25.6 & 7.6 & 7.7 & 2.9 & 233.0 \\
Hill-climbing, repeat 4 & 32.0 & 30.6 & 67.1 & 21.2 & 10.8 & 12.7 & 23.4 & 7.2 & 8.0 & 4.8 & 217.6 \\
Hill-climbing, repeat 5 & 19.9 & 22.6 & 67.4 & 20.8 & 10.5 & 12.8 & 25.6 & 7.3 & 7.7 & 2.9 & 197.5 \\
gpt-6-luna, low & 51.1 & 17.9 & 82.0 & 17.8 & 15.2 & 12.7 & 14.4 & 7.5 & 7.7 & 4.7 & 230.9 \\
gpt-6-luna, high & 58.6 & 19.2 & 96.5 & 32.9 & 13.1 & 13.6 & 25.7 & 7.6 & 8.1 & 8.6 & 284.0 \\
gpt-6-luna, extra-high & 57.8 & 33.8 & 111.4 & 44.2 & 15.4 & 15.5 & 14.4 & 7.6 & 8.1 & 8.6 & 316.7 \\
gpt-6-sol, high & 60.3 & 42.7 & 115.2 & 39.0 & 20.9 & 23.5 & 31.4 & 16.3 & 8.3 & 9.4 & 367.0 \\
gpt-6.1-sol, high & 133.8 & 45.4 & 113.3 & 44.8 & 43.1 & 23.5 & 40.0 & 16.4 & 12.1 & 9.4 & 481.9 \\
\bottomrule
\end{tabular}
\end{table*}

\begin{table*}[!htbp]
\centering
\footnotesize
\setlength{\tabcolsep}{3.5pt}
\caption{As Table~\ref{tab:runcells-dist}, for the CLER objective ($\mathcal{S}_{\mathrm{CLER}}$, Eq.~(3)).  The five random-mutation runs use the random-mutation program of the FOM runs, and the check-weight cap of each child is drawn from 9 to 12.}
\label{tab:runcells-circ}
\begin{tabular}{lrrrrrrrrrrr}
\toprule
Run & 0 & 1 & 2 & 3 & 4 & 5 & 6 & 7 & 8 & 9 & Run score \\
\midrule
Random mutation, repeat 1 & 6.73 & 6.16 & 6.84 & 5.38 & 4.56 & 4.11 & 3.62 & 3.52 & 2.92 & 2.63 & 46.48 \\
Random mutation, repeat 2 & 6.93 & 5.52 & 7.01 & 5.89 & 4.47 & 4.07 & 3.59 & 3.25 & 3.21 & 2.62 & 46.57 \\
Random mutation, repeat 3 & 7.03 & 6.86 & 6.44 & 5.27 & 4.58 & 4.19 & 3.77 & 3.57 & 2.95 & 2.61 & 47.26 \\
Random mutation, repeat 4 & 7.11 & 6.26 & 6.33 & 5.43 & 4.63 & 4.10 & 3.61 & 3.43 & 3.18 & 2.64 & 46.72 \\
Random mutation, repeat 5 & 7.08 & 7.09 & 6.61 & 5.26 & 4.77 & 4.07 & 3.59 & 3.49 & 2.70 & 2.72 & 47.38 \\
Hill-climbing, repeat 1 & 6.70 & 6.83 & 6.70 & 6.13 & 4.82 & 4.82 & 4.13 & 3.51 & 3.42 & 2.68 & 49.75 \\
Hill-climbing, repeat 2 & 7.04 & 6.30 & 7.22 & 6.13 & 4.85 & 4.68 & 3.73 & 3.52 & 2.50 & 2.61 & 48.58 \\
Hill-climbing, repeat 3 & 6.84 & 6.69 & 6.56 & 5.96 & 4.82 & 4.60 & 4.26 & 3.50 & 3.34 & 3.13 & 49.72 \\
Hill-climbing, repeat 4 & 7.04 & 7.11 & 7.03 & 6.03 & 4.79 & 4.60 & 4.54 & 3.65 & 3.10 & 2.62 & 50.52 \\
Hill-climbing, repeat 5 & 7.22 & 7.08 & 7.17 & 6.00 & 5.16 & 4.64 & 4.65 & 3.53 & 3.36 & 2.67 & 51.48 \\
gpt-6-luna, low & 7.24 & 6.69 & 7.02 & 6.77 & 5.04 & 4.53 & 4.06 & 3.15 & 3.31 & 3.22 & 51.03 \\
gpt-6-luna, high & 7.27 & 6.60 & 7.07 & 6.05 & 5.38 & 4.55 & 4.24 & 3.67 & 3.40 & 3.25 & 51.48 \\
gpt-6-luna, extra-high & 7.17 & 6.80 & 6.99 & 6.29 & 4.83 & 4.53 & 4.83 & 3.99 & 3.07 & 3.41 & 51.92 \\
gpt-6-sol, high & 7.36 & 7.40 & 7.43 & 7.11 & 6.48 & 4.96 & 4.87 & 4.12 & 3.35 & 3.48 & 56.57 \\
gpt-6.1-sol, high & 7.34 & 7.36 & 7.34 & 7.08 & 6.16 & 5.21 & 5.06 & 4.21 & 3.35 & 3.52 & 56.62 \\
\bottomrule
\end{tabular}
\end{table*}

\clearpage
\bibliography{zd_ref}

\end{document}